\documentclass[acmsmall, nonacm, authorversion]{acmart}
\usepackage{pythonhighlight}
\usepackage{mathtools}
\usepackage[lined, boxed, ruled, noend]{algorithm2e}
\usepackage[textsize=tiny]{todonotes}
\usepackage{algpseudocode}

\usepackage{listings}
\usepackage{graphicx}
\usepackage{wrapfig}
\usepackage{subcaption}
\usepackage{svg}

\usepackage{xcolor}
\definecolor{mGreen}{rgb}{0,0.6,0}
\definecolor{mGray}{rgb}{0.5,0.5,0.5}
\definecolor{mPurple}{rgb}{0.58,0,0.82}
\definecolor{backgroundColour}{rgb}{0.95,0.95,0.92}

\SetCommentSty{mycommfont}

\AtBeginDocument{%
  \providecommand\BibTeX{{%
    \normalfont B\kern-0.5em{\scshape i\kern-0.25em b}\kern-0.8em\TeX}}}
\usepackage{pifont}
\newcommand{\cmark}{\ding{51}}%
\newcommand{\xmark}{\ding{55}}%
\newcommand{\removethis}[1]{}
\newcommand{\ourtool}{CoNST}

\begin{document}
\title{\ourtool: Code Generator for Sparse Tensor Networks}

\author{Saurabh Raje}
\email{saurabh.raje@utah.edu}
\affiliation{%
  \institution{University of Utah}
  \city{Salt Lake City}
  \state{Utah}
  \country{USA}
}

\author{Yufan Xu}
\email{yf.xu@utah.edu}
\affiliation{%
  \institution{University of Utah}
  \city{Salt Lake City}
  \state{Utah}
  \country{USA}
}

\author{Atanas Rountev}
\email{rountev@cse.ohio-state.edu}
\affiliation{%
  \institution{Ohio State University}
  \city{Columbus}
  \country{USA}
}

\author{Edward F. Valeev}
\email{valeev-76@vt.edu}
\affiliation{%
 \institution{Virginia Tech}
 \city{Blacksburg}
 \country{USA}}

\author{Saday Sadayappan}
\email{saday@cs.utah.edu}
\affiliation{%
  \institution{University of Utah}
  \country{USA}}


\begin{abstract}
Sparse tensor networks are commonly used to represent contractions over sparse tensors. Tensor contractions are higher-order analogs of matrix multiplication. Tensor networks arise commonly in many domains of scientific computing and data science. After a transformation into a tree of binary contractions, the network is implemented as a sequence of individual contractions. Several critical aspects must be considered in the generation of efficient code for a contraction tree, including sparse tensor layout mode order, loop fusion to reduce intermediate tensors, and the interdependence of loop order, mode order, and contraction order.

We propose \ourtool, a novel approach that considers these factors in an integrated manner using a single formulation. Our approach creates a constraint system that encodes these decisions and their interdependence, while aiming to produce reduced-order intermediate tensors via fusion. The constraint system is solved 
by the Z3 SMT solver and the result is used to create the desired fused loop structure and tensor mode layouts for the entire contraction tree. 
This structure is lowered to the IR of the TACO compiler, which is then used to generate executable code. 
Our experimental evaluation demonstrates very significant (sometimes orders of magnitude) performance improvements over current state-of-the-art sparse tensor compiler/library alternatives.



\end{abstract}


\maketitle
\section{Introduction}
\label{sec:intro}

This paper describes \ourtool, a \textbf{co}de generator for \textbf{n}etworks of \textbf{s}parse \textbf{t}ensors. Sparse tensor networks are commonly used 
to represent collections of \emph{tensor contractions} over sparse tensors. Tensor contractions are higher-order analogs of matrix-matrix multiplication. For example, the binary tensor contraction $Y_{ijlm} = U_{ijk} \times W_{klm}$ represents the computation
$$\forall {i,j,l,m}: Y_{ijlm} = \sum \limits_{k} U_{ijk} \times W_{klm} $$

\begin{wrapfigure}{r}{0.25\textwidth}
\begin{center}
\vspace*{-2em}\includegraphics[scale=0.21]{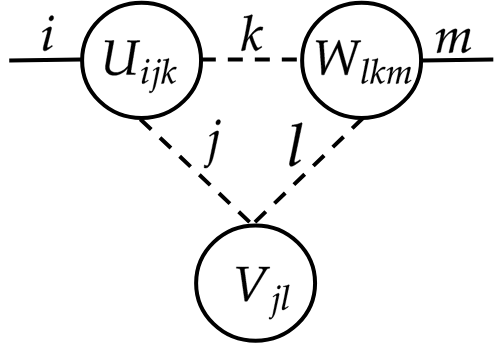}
Tensor network example
\end{center}
\end{wrapfigure}
Multi-tensor product expressions, e.g., $Z_{im} = U_{ijk} \times V_{jl} \times W_{klm}$, arise commonly in many domains of scientific computing and data science
(e.g., high-order models in quantum chemistry \cite{riplinger2016sparse}, tensor decomposition schemes \cite{kolda2009tensor}).
Such expressions involve multiple tensors and multiple summation indices: 
$$ \forall {i,m}: Z_{im} = \sum \limits_{j,k,l } U_{ijk} \times V_{jl} \times W_{klm} $$
Such multi-tensor products are also referred to as \emph{tensor networks}, represented with a node for every tensor instance and edges representing the indices that index the various tensors. 
Efficient evaluation of such an expression typically requires a transformation into a tree of binary contractions, which is then executed as a sequence of these individual contractions. 



Many efforts have 
been directed towards compiler optimization of sparse matrix and tensor computations \cite{kjolstad2017tensor,workspaces2019cgo,strout2018sparse,liu2021sparta, liu2021athena,tian2021high,dias2022sparselnr,zhao2022polyhedral,cheshmi2023runtime}.
However, the current state of the art does not adequately address a number of critical inter-dependent aspects in the generation of efficient code for 
a given tree of sparse binary contractions. 


\noindent \textbf{Sparse tensor layout mode order:} The most commonly used representation for efficient sparse tensor computations is the CSF (Compressed Sparse Fiber) format \cite{smith-csf}, detailed in Sec.~\ref{sec:background}. Since CSF uses a nested representation with $n$ levels for a tensor of order $n$, efficient access is only feasible for some groupings of non-zero elements by traversing the hierarchical nesting structure. Selecting the order of the $n$ modes of a tensor is a key factor for achieving high performance. 
Prior efforts in compiler optimization and code generation for sparse computations 
have not explored the impact of the choice of the CSF layout mode order on the performance of contraction tree evaluation. 

\noindent \textbf{Loop fusion to reduce intermediate tensors:} 
The temporary intermediate tensors that correspond to inner nodes of the contraction tree could be 
much larger than the input and output tensors of the network. 
By fusing common loops in the nested loops implementing the contractions that produce and consume an intermediate tensor, the size of that tensor can be reduced significantly (as illustrated by an example in Sec.~\ref{sec:background}). 

\noindent \textbf{Inter-dependence between loop order, mode order, and contraction order:} In addition to selecting the layout mode order for each tensor in the contraction tree, code generation needs to  select a legal loop fusion structure to implement the contractions from the tree. 
Such a fused structure depends on the order of surrounding loops for each contractions, on the order in which the contractions are executed, and on the choice of layout mode order. No existing work considers the space of these inter-related choices in a systematic and general manner.


\noindent \textbf{Our solution:} We propose \ourtool, a novel approach that considers these factors in an integrated manner using a single formulation. Our approach creates a constraint system that encodes these decisions and their interdependence, while aiming to produce reduced-order intermediate tensors via fusion. The constraint system is solved 
by the Z3 SMT solver \cite{z3} and the result is used to create the desired fused loop structure and tensor mode layouts for the entire contraction tree. 
This structure is lowered to the IR of the TACO compiler \cite{kjolstad2017tensor}, which is then used to generate the final executable code. The main contributions of \ourtool\ are as follows:
\begin{itemize}
    \item We design a novel constraint-based approach for encoding the space of possible fused loop structures and tensor CSF layouts, with the goal of reducing the order of intermediate tensors. This is the first work that proposes such a general integrated view of code generation for sparse tensor contraction trees.
    \item We develop an approach to translate the constraint solution to the concrete index notation IR \cite{workspaces2019cgo} of the TACO compiler. 
   \item We perform extensive experimental comparison with the three most closely related systems: TACO~\cite{kjolstad2017tensor}, SparseLNR~\cite{dias2022sparselnr}, and Sparta~\cite{liu2021sparta}. Using a variety of benchmarks from quantum chemistry and tensor decomposition, we demonstrate very significant (sometimes orders of magnitude) performance improvements over this state of the art.
\end{itemize}

\section{Background and Overview}
\label{sec:background}
\subsection{Tensor Networks}

We first describe the abstract specification of a tensor network. Such a specification can be lowered to many possible code implementations. Examples of such implementations are also given below.

\paragraph{Sparse tensors} A tensor $T$ of order $n$ is defined by a sequence $\langle d_0, \ldots, d_{n-1} \rangle$ of \emph{modes}. Each mode $d_k$ denotes a set of index values: $d_k = \{ x \in \mathbb{N} : 0\le x < N_k\}$, where $N_k$ is the \emph{mode extent}. Note that the numbering of modes from $0$ to $n-1$ is purely for notational purposes and does not imply any particular concrete data layout representation; deciding on such a layout is one of the goals of our work, as described later.


For a sparse tensor $T$, its non-zero structure is defined by some subset $\mathit{nz}(T)$ of the Cartesian product $d_0\times d_2 \times \ldots \times d_{n-1} $. All and only non-zero elements of $T$ have coordinates that are in $\mathit{nz}(T)$. Each $(x_0,x_1,\ldots) \in \mathit{nz}(T)$ is associated with a non-zero value $T(x_0,x_1,\ldots)\in \mathbb{R}$. 

\paragraph{Tensor references} The tensor expressions described below use \emph{tensor references}. For each tensor used in the computation, there may be one or more references in these expressions. A reference to an order-$n$ tensor $T$ is defined by a sequence $\langle i_0, \ldots, i_{n-1} \rangle$ of distinct \emph{iteration indices} (``indices'' for short). 
Such a reference will be denoted by $T_{i_0 i_1 \ldots}$. Each index $i_k$ is mapped to the corresponding mode $d_k$ of $T$ and denotes the set of values defined by that mode: $i_k = \{ x \in \mathbb{N} : 0\le x < N_k\}$.

The same index may appear in several tensor references, for the same tensor or for different ones. In all such occurrences, the index denotes the same set of index values. For example, an expression discussed shortly contains tensor references $ X_{ i j q r}$, $A_{ i p q }$, and $B_{ j p r }$.
As an illustration, index $j$ appears in two of these references, and is mapped to mode $1$ of $X$ and mode $0$ of $B$ (and thus both modes have the same extent). 

\paragraph{CSF representation}
As discussed earlier, our work focuses on sparse tensors represented in the CSF (Compressed Sparse Fiber) \cite{smith-csf} format.
CSF organizes a sparse tensor as a tree, defined by some permutation of modes $d_0, \ldots, d_{n-1}$. This order of modes defines the CSF layout and must be decided when creating a concrete implementation of a computation that uses the tensor. 
The internal nodes of the tree store the indices of non-zero elements in the corresponding mode. 
The leaves of the tree store the non-zero values. An auxiliary root node connects the entire structure. 

\begin{figure}[t]
    \includegraphics[scale=0.25]{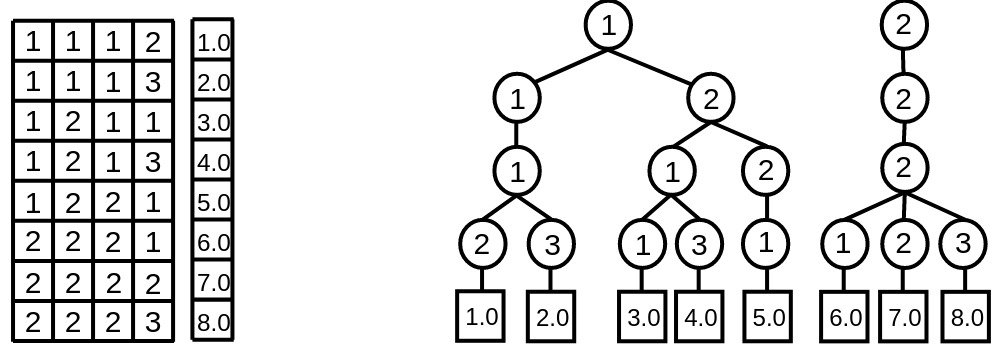}
    \caption{\label{fig:csf-example}The CSF format for representing an order-$4$ sparse tensor in memory. The table on the left shows the indices of non-zero elements. The tree on the right shows the CSF representation (root node is not shown). 
   }
\end{figure}

Fig.~\ref{fig:csf-example} illustrates the CSF representation for an order-$4$ sparse tensor.
When the abstract specification of a tensor expression (or equivalently, of a tensor network) is lowered to a concrete implementation, both tensors and tensor references are instantiated to specific representations. For example, suppose we have a tensor $A$ with modes $d_0$, $d_1$, and $d_2$, and a reference $A_{i p q}$ appears in the tensor network. One (of many) possible implementations is to order the modes as $d_1$, $d_2$, $d_0$ in outer-to-inner CSF order. The code references to the tensor would be consistent with this order, i.e., reference $A_{i p q}$  becomes \texttt{A[p,q,i]} in the code implementation. A fundamental question is how to select the order of modes for each tensor in the network. The constraint-based approach described in the next section encodes all possible orders by employing constraint variables.

\paragraph{Tensor contractions} Consider tensors $T$, $S$, and $R$.
A \emph{binary contraction} $R_{i_0 i_1 \ldots} = T_{j_0 j_1 \ldots} \times S_{k_0 k_1\ldots}$ contains three tensors references. 
Let $I_T$, $I_S$, and $I_R$ denote the sets of indices appearing in each reference, respectively. 
The contraction has the following properties:
\begin{itemize}
\item The non-zero structure of the result $R$ is defined by the non-zero structure of $T$ and $S$ as follows. 
A tuple $(z_0,z_1,\ldots)$ is in $\mathit{nz}(R)$ if and only if there exists at least one pair of tuples 
$(x_0,x_1,\ldots) \in \mathit{nz}(T)$ and $(y_0,y_1,\ldots) \in \mathit{nz}(S)$ such that for each index $i \in I_T \cup I_S \cup I_R$, the values corresponding to $i$ in the three tuples (if present) are the same. 
\item For any $(z_0,z_1,\ldots) \in \mathit{nz}(R)$, the associated value $R(z_0,z_1,\ldots)\in \mathbb{R}$ is the sum of
$T(x_0,x_1,\ldots) \times S(y_0,y_1,\ldots)$ for all such pairs of tuples $(x_0,x_1,\ldots) \in \mathit{nz}(T)$ and $(y_0,y_1,\ldots) \in \mathit{nz}(S)$.
\end{itemize}

As a simple example, $R_{i j} = T_{i k} \times S_{k j}$ represents a standard matrix multiplication: for any $(a,b) \in \mathit{nz}(R)$ we have $R(a,b) = \sum_{\{c : (a, c) \in \mathit{nz}(T) \wedge (c,b) \in \mathit{nz}(S) \}} T(a,c) \times S(c,b)$.


The indices from $I_T \cup I_S \cup I_R$ can be classified into two categories. Any index $i\in I_R$ is an \emph{external} index for this contraction. Any index $i\in I_C = (I_T\cup I_S)\setminus I_R$ is a \emph{contraction} index for the contraction. 

\begin{figure}
\begin{minipage}{.45\textwidth}
    \begin{center} 
    \includegraphics[scale=0.2]{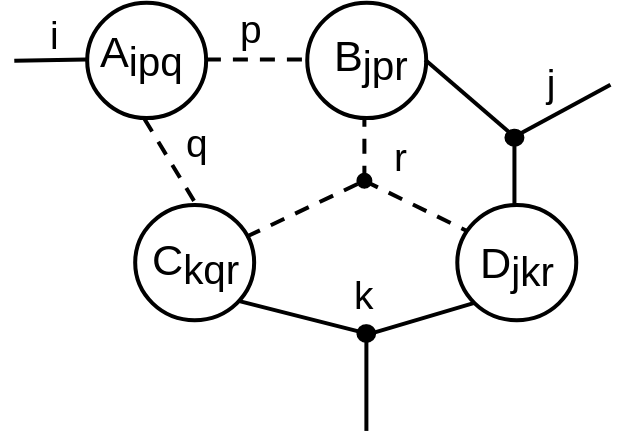}\\
    (a) Tensor network
    \end{center}
\end{minipage}
\hfill
\begin{minipage}{.50\textwidth}
{\small
\begin{lstlisting}
R[*,*,*] = 0
for i in [0,Ni)
 for j in [0,Nj)
  for k in [0,Nk)
   for p in [0,Np)
    for q in [0,Nq)
     for r in [0,Nr)
      R[i,j,k] += A[i,p,q]*B[j,p,r]*
                  C[k,q,r]*D[j,k,r]
\end{lstlisting}
}
    \begin{center}
        (b) Direct $n$-ary contraction
    \end{center}
\end{minipage}

\caption{Tensor network and code for direct $n$-ary contraction for expression $R_{i j k} = A_{ i p q } \times B_{ j p r } \times C_{ k q r } \times D_{ j k r } $}
\label{fig:example_tensor_network}
\end{figure}

\paragraph{Tensor networks} 
The meaning of a general (non-binary) contraction expression of the form $R_{\ldots} = \mathit{T1}_{\ldots} \times \ldots \times \mathit{Tn}_{\ldots} $ is defined similarly. A general tensor contraction expression comprised of a tensor product of many tensor references can be equivalently represented as a \emph{tensor network}, with one vertex for each tensor reference in the expression, and a hyper-edge for every index. An example of a tensor network representing the tensor expression $R_{i j k} = A_{ i p q } \times B_{ j p r } \times C_{ k q r } \times D_{ j k r } $ is shown in Fig.~\ref{fig:example_tensor_network}(a). Here dashed hyperedges are used to distinguish the \emph{contraction} indices in the tensor expression (i.e., $i$, $j$, and $k$) from the \emph{external} indices. Note that the result tensor is not explicitly represented in this network, but rather implicitly defined by the external indices.

The direct computation of any tensor network (multi-tensor product expression) can be performed via a nested loop computation, with one loop corresponding to each index, and a single statement that mirrors the tensor expression. Fig.~\ref{fig:example_tensor_network}(b) illustrates this approach. Note that the figure shows a specific code version with a concrete loop order (e.g., $i$ in the outermost position) and tensor data layouts (e.g., $j$ is the outermost CSF level of tensor $D$). There are many possible choices for the loop order and the tensor layout, as elaborated later.

\begin{figure}
\begin{minipage}{.35\textwidth}
    \begin{center}\includegraphics[scale=0.2]{sptens_network.png}\\
\hspace*{0em}(a) Tensor network
    \end{center}
\end{minipage}
\begin{minipage}{.35\textwidth}
    \begin{center}
    \hspace*{-5.5em}\includegraphics[scale=0.2]{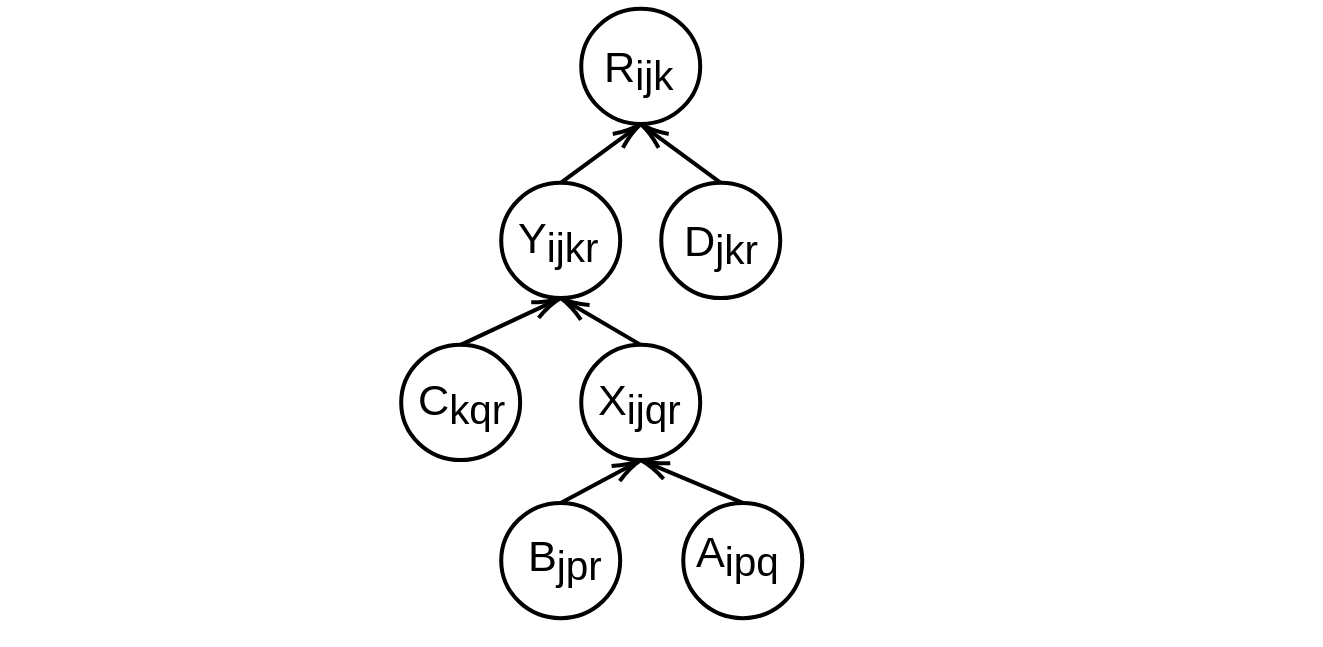}\\
\hspace*{0em}(b) Contraction tree
    \end{center}
\end{minipage}
\hfill
\begin{minipage}{.25\textwidth}
    \begin{center}
$ X_{ i j q r} = A_{ i p q } \times B_{ j p r }$\\
$ Y_{ i j k r} = X_{ i j q r } \times C_{ k q r } $ \\ 
$ R_{ i j k } = Y_{i j k r } \times D_{ j k r } $ \\
~ \\
        (c) Binary contractions
    \end{center}
\end{minipage}
\caption{Contraction tree for a tensor network\label{fig:synth_dag}}
\end{figure}

\paragraph{Contraction tree}
The computational complexity of such an implementation is $\mathcal{O}(N_i N_j N_k N_p N_q N_r)$. However, by exploiting associativity and distributivity, the multi-term product can be rewritten as a sequence of binary contractions, with temporary intermediate tensors $X$ and $Y$ as shown in Fig.~\ref{fig:synth_dag}(c).
By using a \emph{sequence of binary contractions} instead of a direct $n$-ary contraction, the complexity is significantly reduced to $\mathcal{O}(N_i N_j N_p N_q N_r + N_i N_j N_k N_q N_r + N_i N_j N_k N_r)$. If all tensor modes have the same extent $N$, the complexity reduces from $\mathcal{O}(N^6)$ 
to $\mathcal{O}(N^5)$.

In general, there exist many different sequences of binary tensor contractions to compute a tensor network, with varying computational complexity. The problem of identifying an operation-optimal sequence of binary contractions for a multi-term product expression has been extensively studied \cite{kanakagiri2023minimum}. In this paper, we do not address this issue, and assume that an operation-optimal binarization of a tensor network has already been performed and provided as the input to our code generator. Such a binarization can be expressed as a \emph{tensor contraction tree}, illustrated in Fig.~\ref{fig:synth_dag}(b). For readability, Fig.~\ref{fig:synth_dag}(a) repeats the original tensor network described earlier. 

\subsection{Challenges and Overview of Solution}
The problem we address in this paper is the following: \emph{Given a binary tensor contraction tree for a sparse tensor network, generate efficient code for its evaluation.} While it is straightforward to generate loop code for a sequence of dense tensor contractions, and efficient tensor contraction libraries are also available \cite{cuTensor, hirata2003tensor}, it is not so for sparse tensors. Even with dense tensors, although loop code is easy to generate, a problem arises if the size of intermediate tensors is too large. For our example, if all tensor modes have the same extent $N$, the intermediate tensors $X$ and $Y$ require $N^4$ space versus $N^3$ for the input and output tensors. Thus, the intermediate tensors may be too large to fit in available memory. However, by using loop fusion, the required sizes of intermediate tensors can be significantly reduced. Fig.~\ref{fig:fusion}(a) shows one possible code implementation for the contraction tree from Fig.~\ref{fig:synth_dag}(b). Since identical loops over indices $i$ and $j$ exist in the loop code for all three binary contractions, we can fuse those loops to create the imperfectly nested loop structure shown in Fig.~\ref{fig:fusion}(b), where the space required for the intermediate tensors has been reduced from $N^4$ to $N^2$.
\begin{figure}[!ht]
\begin{minipage}{.45\textwidth}
{\scriptsize
\begin{lstlisting}
    X[*,*,*,*] = 0 
    Y[*,*,*,*] =0
    R[*,*,*] = 0
    for i,j,p,q,r
       X[i,j,q,r] += A[i,p,q]*B[j,p,r]
    for i,j,k,q,r
       Y[i,j,k,r] += X[i,j,q,r]*C[k,q,r]
    for i,j,k,r
       R[i,j,k] += Y[i,j,k,r]*D[j,k,r]
\end{lstlisting}
 \label{lst:unfused_pseudocode}
}
\begin{center}
  (a) Unfused sequence of contractions
\end{center}
\end{minipage}
\hfill
\begin{minipage}{.45\textwidth}
{\scriptsize
\begin{lstlisting}
    R[*,*,*]=0
    for i,j
      X[*,*]=0
      Y[*,*]=0 
      for p,q,r
         X[q,r] += A[i,p,q]*B[j,p,r]
      for k,q,r
         Y[k,r] += X[q,r]*C[k,q,r]
      for k,r
         R[i,j,k] += Y[k,r]*D[j,k,r]
\end{lstlisting}
}
\begin{center}
  (b) Common loops $i$ and $j$ fused
\end{center}
\end{minipage}
    \caption{Reduction of size of intermediate tensors via loop fusion}
    \label{fig:fusion}
\end{figure}


For dense tensors, the main benefit of loop fusion for a sequence of tensor contractions is the reduction of sizes of temporary intermediate tensors. This reduction is useful primarily when intermediate tensors are too large to fit in main memory (or in global memory, for GPU execution). This is because loop tiling can be used very effectively to achieve very high operational intensities, whether fusion is used or not, i.e., loop fusion does not typically enable significant reductions in data access overheads for a sequence of dense tensor contractions. 

However, \emph{for a sequence of sparse tensor contractions, loop fusion can enable significant performance improvement over unfused execution}.
In contrast to the dense case, with sparse tensor contractions, a fundamental challenge is that of insertion of each additive contribution from the product of a pair of elements of the input tensors to the appropriate element of a sparse output tensor. The TACO compiler \cite{kjolstad2017tensor} defines a \emph{workspaces} optimization \cite{workspaces2019cgo} to address this challenge, where a dense multidimensional temporary array is used to assemble multidimensional slices of the output tensor during the contraction of sparse input tensors. By using a dense ``workspace'', very efficient $\mathcal{O}(1)$ cost access to arbitrary elements in the slice is achieved for assembling the irregularly scattered contributions generated during the contraction. A significant consideration with the use of the dense workspaces is the space required: the extents of the workspace array must equal the extents of the corresponding modes of the sparse output tensor and thus can become excessive. By use of loop fusion between producer and consumer contractions to reduce the number of explicitly represented modes in intermediate tensors, we can make efficient use of TACO's workspaces optimization.

In addition to fusion, a critical factor for high performance is the compatibility between loop order and layout order. 
For sparse tensors represented in CSF format, efficient access to the non-zero elements is only feasible if the outer-to-inner order of nested loop indices in the code implementation is consistent with the layout order of tensor modes, in relation to the loop indices that index them. For example, the elements referenced by \texttt{A[i,p,q]} can be accessed efficiently only if $i$ appears earlier than $p$ (which itself appears earlier than $q$) in the loops surrounding this reference.  

Given a binary contraction tree to implement a general sparse tensor expression, three critical inter-related decisions affect the achieved performance of the generated code:
\begin{itemize}
    \item {\bf Linear execution order of contractions:} The fusibility of loops between a \emph{producer} contraction of an intermediate tensor and a subsequent \emph{consumer} contraction is affected by the linear execution order of the contractions.
    \item {\bf Loop permutation order for each  contraction:} All surrounding loops of a tensor contraction are fully permutable. The chosen permutation affects both the fusibility of loops across tensor contractions as well as the efficiency of access of the non-zero elements of sparse tensors in the contraction.
    \item {\bf Mode layout order for each tensor:} The compatibility of the layout order of each tensor with the loop order of the surrounding loops is essential for efficient access. 
\end{itemize}

These three decisions are inter-dependent. The linear execution order (i.e., the topological sort of the contraction tree) affects which loop fusion structures are possible. The order of loops for each contraction determines what fusion can be achieved, while also imposing constraints on the data layouts of tensors that appear in the contraction tree. In this paper, we propose a novel integrated solution that considers these three decisions in a single formulation. Our approach creates a constraint system that encodes the space of possible decisions and their interdependence. This system is then solved using the Z3 SMT solver \cite{z3}. The solution is used to create a legal fused loop structure that reduces the size of intermediate tensors while ensuring the compatibility constraints described above. To the best of our knowledge, this is the first work that takes such an integrated view and provides a general approach for code generation for arbitrary tensor contraction trees. Table~\ref{tab:comparison_SOA} contrasts our work with the three most closely related state-of-the-art systems for sparse tensor computations, discussed below.



\noindent \textbf{TACO} \cite{kjolstad2017tensor}: The \ourtool\ system leverages, as its last stage, the 
code generator for sparse tensor 
    computations in the Tensor Algebra Compiler (TACO). The main focus of the TACO framework is the generation of efficient code for $n$-ary contractions with arbitrarily complex tensor expressions. 
    While TACO can be used to generate code for a sequence of binary sparse tensor contractions, it does not address optimizations like loop fusion across tensor contractions, tensor mode layout choice, or the choice of sequence of tensor contractions for a given contraction tree. In our experimental evaluation (Sec.~\ref{sec:exp}), we show that code generated by \ourtool\ achieves significant speedup over code directly generated by TACO.

    
    \noindent \textbf{SparseLNR} \cite{dias2022sparselnr} builds on TACO to implement loop fusion optimization. It takes a multi-term tensor product expression as input and generates fused loop code for a sequence of binary tensor contractions corresponding to the input tensor product expression.
    In our experimental evaluation, we compare the performance of code generated by SparseLNR with code generated by \ourtool\ and demonstrate significant speedups.
    
    
    \noindent 
    \textbf{Sparta}~\cite{liu2021sparta} implements a library for efficient tensor contraction of arbitrary pairs of sparse tensors. Since it is a library function, it does not address any optimizations like loop fusion across contractions, data layout choice for tensors, or the schedule of contractions for a contraction tree. We performed extensive experimentation to compare the performance of code generated by \ourtool\ with the best performance among all valid tensor layout permutations for unfused sequences of contractions executed using Sparta. These experiments demonstrate very significant performance gains for \ourtool. 

\begin{table}
    \centering
        \caption{Comparison with state-of-the-art systems for sparse tensor computations\label{tab:comparison_SOA}}
    \begin{tabular}{|l|c|c|c|c|} \hline
         &  TACO &  SparseLNR &  Sparta & \ourtool\ (ours)\\ \hline
         Loop fusion &  {\color{red}{\color{red}\xmark}} &  {\color{green}\cmark} &  {\color{red}\xmark} & {\color{green}\cmark}\\ \hline
         Data layout selection &  {\color{red}\xmark} &  {\color{red}\xmark} &  {\color{red}\xmark} & {\color{green}\cmark}\\ \hline
         Schedule for contraction trees & {\color{red}\xmark} & {\color{red}\xmark} & {\color{red}\xmark} & {\color{green}\cmark} \\\hline
    \end{tabular}
\end{table}
\section{Constraint-Based Integrated Fusion and Data Layout Selection}
\label{sec:constraints}

Our approach aims to generate a concrete implementation of a given contraction tree by automatically determining (1) the order of modes in the data layout of each tensor, and (2) a structure of fused loops that minimizes the order of intermediate tensors. We formulate a constraint system that answers the following question: For the given contraction tree, does there exist an implementation for which all intermediate tensors are of order at most $l$, for some given integer $l$? We first ask this question for $l=1$. If the answer is positive, the constraint system solution is used to construct a code implementation for the contraction tree.
If the answer is negative, we formulate and solve a constraint system for $l=2$, seeking a solution in which all intermediates are at most 2D matrices. This process continues until we find a solution. Note that a trivial solution without any fusion is guaranteed to exist for a sufficiently large value of $l$.

In each of these steps, we employ the Z3 SMT solver \cite{z3} to provide either (1) a negative answer (``the constraint system is unsatisfiable''), or (2) a positive answer with a concrete constraint solution that defines the desired tensor layouts and loop structure.  

\subsection{Input and Output}


The input to our approach is a set of contractions $\{ C_0, C_1, \ldots, C_{m-1} \}$ organized in a contraction tree. Each leaf node corresponds to an input tensor reference, the root node corresponds to a result tensor reference, and every other node corresponds to an intermediate tensor reference. As an example, the contraction tree for $ X_{ i j q r} = A_{ i p q } \times B_{ j p r }; Y_{ i j k r} = X_{ i j q r } \times C_{ k q r }; R_{ i j k } = Y_{i j k r } \times D_{ j k r } $ 
was shown earlier in Fig.~\ref{fig:synth_dag}(b). Here $A$, $B$, and $C$ are input tensors, $X$ and $Y$ are intermediate tensors, and $R$ is the result tensor. 


A naive implementation of a given tree would contain a sequence of perfectly nested loops (one loop nest per contraction), based on some valid topological sort order of tree nodes. For each contraction, the loop nest would be some permutation of the set of indices that appear in the tensor references, and the loop body would be a single assignment. For example, the loop nest for
$ X_{ i j q r} = A_{ i p q } \times B_{ j p r }$
would contain loops for $r$, $q$, $i$, $j$, and $p$ in some order. 

As discussed earlier in Section~\ref{sec:background}, for any (unfused or fused) implementation, 
a fundamental constraint is that the order of surrounding loops must match the data layout order of modes in the CSF tensor representation. This is needed to allow for efficient iteration over the sparse representation. For example, consider reference $A_{ i p q }$. Recall from the earlier discussion that each index is mapped to the corresponding mode of $A$: $i$ is mapped to $d_0$, $p$ is mapped to $d_1$, and $q$ is mapped to $d_2$. A concrete implementation would select a particular order of $d_0$, $d_1$, and $d_2$ as the outer, middle, and inner level in the CSF representation. For example, suppose that this order is, from outer to inner, $\langle d_1, d_2, d_0 \rangle$. In the code implementation, the tensor reference would be \texttt{A[p,q,i]}. Efficient iteration over elements of $A$ would require that the loop structure surrounding the reference matches this order: the $p$ loop must appear before the $q$ loop, which must appear before the $i$ loop. The constraint-based approach described below incorporates such constraints for the loops that surround (in a fused code structure) each tensor reference from the contraction tree.

Each of the fused loop structures we would like to explore can be uniquely defined by (1) a topological sort order of the non-leaf nodes in the contraction tree, and (2) 
for each such node, an ordering of the indices that appear in it. The index order for a node defines the order of loops that would surround the corresponding assignment in the fused loop nest. This order also defines the CSF layout order for the corresponding tensors.

\begin{wrapfigure}{l}{0.45\textwidth}
  \begin{center}
\vspace*{-1em}    {\small
\begin{lstlisting}
for r,j
  for p,q,i
     X[q,i] += A[p,q,i]*B[r,j,p]
  for q,k,i
     Y[k,i] += X[q,i]*C[r,q,k]
  for k,i
     R[j,k,i] += Y[k,i]*D[r,j,k]
\end{lstlisting}
}
  \end{center}
  \caption{Fused code structure\label{fig:fused}}
\end{wrapfigure}
For example, consider the following code structure, which is derived from the solution of our constraint system for the running example. Here there is a single valid topological sort for the assignments. The ordering of surrounding loops for the  assignments is $\langle r, j, p, q,i \rangle$,
$\langle r, j, q, k, i \rangle$ and $\langle r, j, k, i\rangle$, respectively. The fusion of the common $r$ and $j$ loops allows $X$ and $Y$ to be reduced to 2D tensors. The order of indices in all tensor references is consistent with the order of surrounding loops. 

\subsection{Constraint Formulation}
The space of targeted code structures is encoded via constraints over integer-typed constraint variables. The following constraint variables and corresponding constraints are employed. 

\subsubsection{Ordering of assignments}
First, for each contraction $C_i$, the position of the corresponding assignment relative to the other assignments in the code is encoded by a constraint variable $\mathit{ap}_i$ (short for ``assignment position for $C_i$'') such that 
\begin{description}
    \item[] $0 \le \mathit{ap}_i < m$
    \item[] $ \mathit{ap}_i \ne \mathit{ap}_k$ for all $k\ne i$
    \item[] $ \mathit{ap}_i < \mathit{ap}_j$ if $C_i$ is a child of $C_j$ in the contraction tree    
\end{description}
Here $m$ is the number of contractions. 
The first two constraints guarantee uniqueness and appropriate range for all $\mathit{ap}_i$.
The last constraint ensures a valid topological sort order. Any variable values that satisfy these constraints define a particular valid relative order for the corresponding assignments. For the running example, we have $\mathit{ap}_0$ for $ X_{ i j q r} = A_{ i p q } \times B_{ j p r }$, 
$\mathit{ap}_1$ for $Y_{ i j k r} = X_{ i j q r } \times C_{ k q r }$, and $\mathit{ap}_2$ for $R_{ i j k } = Y_{i j k r } \times D_{ j k r } $. For this particular contraction tree the only possible solution is $\mathit{ap}_i=i$. In a more general tree, there may be multiple valid assignments of values to $\mathit{ap}_i$, each corresponding to one of the topological sort orders. 

\subsubsection{Ordering of tensor modes}
For each order-$n$ tensor $T$ that has references in the contraction tree, and each mode $d_j$ of $T$ ($0\le j < n$), we use a constraint variable $\mathit{dp}_{T,j}$ to encode the position of $d_j$ in the CSF layout of the tensor. The following constraints are used:  
\begin{description}
    \item[] $0 \le \mathit{dp}_{T,j} < n$
    \item[] $ \mathit{dp}_{T,j} \ne \mathit{dp}_{T,j'}$ for all $j'\ne j$
\end{description}
Any constraint variable values that satisfy these constraints define a particular permutation of the modes of tensor $T$ and thus a concrete CSF data layout for that tensor. 

\textbf{Example.} In the running example $A$ has three modes and thus three constraint variables $\mathit{dp}_{A,0}$, $\mathit{dp}_{A,1}$, and $\mathit{dp}_{A,2}$. In the code structure shown in Fig.~\ref{fig:fused}, abstract tensor reference $A_{ i p q }$ is mapped to concrete reference \texttt{A[p,q,i]}. This corresponds to the following assignment of values to the constraint variables: $\mathit{dp}_{A,0}=2$, $\mathit{dp}_{A,1}=0$, and $\mathit{dp}_{A,2}=1$. Thus, the outermost level in the CSF representation corresponds to mode $d_1$ (indexed by $p$), the next CSF level corresponds to $d_2$ (indexed by $q$), and the inner CSF level corresponds to $d_0$ (indexed by $i$). $\blacksquare$

\subsubsection{Ordering of loops}
Next, we consider constraints that encode the fused loop structure. 
For any contraction $C_i$, we need to encode the loop order of the loops surrounding the corresponding assignment. 
Let $I_i$ be the set of indices that appear in $C_i$. 



For each $k \in I_i$, 
we define an integer constraint variable $\mathit{lp}_{i,k}$ 
(short for ``loop position of index $k$ for $C_i$'').
These variables will encode a permutation of the elements of 
$I_i$---that is,
a loop order for the loops surrounding the assignment for $C_i$. If $\mathit{lp}_{i,k}$ has a value of $0$, 
index $k$ will be the outermost loop surrounding the assignment. If the value is $1$, the index will be the second-outermost loop, etc. To encode a permutation, for each 
$k\in I_i$
we have constraints
\begin{description}
    \item[] $0 \le \mathit{lp}_{i,k} < |I_i|$
    \item[] $ \mathit{lp}_{i,k} \ne \mathit{lp}_{i,k'}$ for all $k' \in I_i \setminus \{ k \} $
\end{description}

\textbf{Example.} In the running example, 
for contraction  $ C_0: X_{ i j q r} = A_{ i p q } \times B_{ j p r }$ we have 
$I_0 = \{ i, j, p, q, r\}$.
For this contraction we will use constraint variables 
$\mathit{lp}_{0,i}$, $\mathit{lp}_{0,j}$, $\mathit{lp}_{0,p}$, $\mathit{lp}_{0,q}$, $\mathit{lp}_{0,r}$. 
In the code structure shown in Fig.~\ref{fig:fused}, the loop order for $C_0$ is $\langle r, j, p, q,i \rangle$. This order corresponds to a constraint solution in which $\mathit{lp}_{0,i}=4$, $\mathit{lp}_{0,j}=1$, $\mathit{lp}_{0,p}=2$, $\mathit{lp}_{0,q}=3$, and $\mathit{lp}_{0,r}=0$. $\blacksquare$



\subsubsection{Consistency between mode order and loop order}
Next, we need to ensure that the order of loops defined by $\mathit{lp}_{i,k}$ is consistent with the order of modes for each tensor appearing in contraction $C_i$, as encoded by $\mathit{dp}_{T,j}$. Consider a reference to $T$ appearing in contraction $C_i$. For each pair of modes $d_j$ and $d_{j'}$ of $T$, let $k$ and $k'$ be the 
indices that correspond to these modes in the reference.
The following constraint enforces the consistency between mode order and loop order:
$$(\mathit{dp}_{T,j} < \mathit{dp}_{T,j'}) \implies (\mathit{lp}_{i,k} < \mathit{lp}_{i,k'})$$
Here $\mathit{dp}_{T,j} < \mathit{dp}_{T,j'}$ is true if and only if mode $d_j$ appears earlier than mode $d_{j'}$ in the concrete CSF data layout of tensor $T$. If this is the case, we want to enforce that the index corresponding to $d_j$ (i.e., $k$)
appears earlier that the index corresponding to $d_{j'}$ (i.e., $k'$)
in the loop order of loops surrounding the assignment for $C_i$. As discussed earlier, this constraint ensures that the order of iteration defined by the loop order allows an efficient traversal of the CSF data structure for $T$. For intermediates that can be implemented with dense workspaces, such constraints are not necessary. 

\textbf{Example.} Consider reference $A_{ i p q }$ from the running example and the pair of modes $d_0$ and $d_2$, with corresponding indices $i$ and $q$. The relationship between variables $\mathit{dp}_{A,0}$ (for $d_0)$, $\mathit{dp}_{A,2}$ (for $d_2)$, 
$\mathit{lp}_{0,i}$ (for $i$), and $\mathit{lp}_{0,q}$ (for $q$) 
is captured by the following two constraints:
$$ (\mathit{dp}_{A,0} < \mathit{dp}_{A,2}) \implies (\mathit{lp}_{0,i} < \mathit{lp}_{0,q}) \hspace*{2em} (\mathit{dp}_{A,2} < \mathit{dp}_{A,0}) \implies (\mathit{lp}_{0,q} < \mathit{lp}_{0,i})$$
As described earlier, in the constraint solution we have 
$\mathit{dp}_{A,0}=2$, $\mathit{dp}_{A,2}=1$, $\mathit{lp}_{0,i}=4$, and $\mathit{lp}_{0,q}=3$.
Of course, these values satisfy both constraints. $\blacksquare$

\subsubsection{Producer-consumer pairs}
Finally, we consider every pair of contractions $C_i, C_j$ such that $C_i$ is a child of $C_j$ in the contraction tree. In this case $C_i$ produces a reference to a tensor $T$ that is then consumed by $C_j$. Let $n$ be the order of $T$. Our goal is to identify a loop fusion structure that reduces the order of this intermediate tensor $T$ to be some $n'\le l$ for a given parameter $l$. Recall that in our overall scheme, we first define a constraint system with $l=1$. If this system cannot be satisfied, we define a new system with $l=2$, etc.

Let $I_T$ be the set of 
indices that appear in the reference to $T$. We define constraints that include $\mathit{lp}_{i,k}$ (for the producer $C_i$) and $\mathit{lp}_{j,k}$ (for the consumer $C_j$), for all $k\in I_T$.
The constraints ensure that a valid fusion structure exists to achieve the desired reduced order $n'$ of $T$. 

\paragraph{Producer constraints} First, we consider the outermost $n-l$ indices in the loop order associated with the producer $C_i$ and ensure that they are all indices
of the result reference. Specifically, for each $s$ such that $0 \le s < n-l$ and for each $k\in I_T$,
we create a disjunction of terms of the form $\mathit{lp}_{i,k}=s$. This guarantees that the loop at position $s$ in the loop structure surrounding the producer statement is iterating over one of the indices that appear in the result reference. The combination of these constraints for all pairs of $s$ and $k$ ensures that the outermost $n-l$ loops for $C_i$ are all indices of its result tensor reference. 

\textbf{Example.} Consider reference $X_{ i j q r }$ from the running example. This reference is produced by $ C_0: X_{ i j q r} = A_{ i p q } \times B_{ j p r }$ and consumed by $C_1: Y_{ i j k r} = X_{ i j q r } \times C_{ k q r }$. We have $I_X = \{ i, j, q, r \}$.
The producer constraints will involve variables 
$\mathit{lp}_{0,i}$, $\mathit{lp}_{0,j}$, $\mathit{lp}_{0,q}$, and $\mathit{lp}_{0,r}$. 

Suppose $l=2$. We would like the outermost $n-l=4-2$ indices in the loop order for $C_0$ to be indices that access this reference. Together with the remaining constraints described shortly, this would allow those two indices to be removed from the reference after fusion. As a result, the order of $X$ can be reduced from $4$ to $2$. Two constraints are formulated. First,
$$\mathit{lp}_{0,i} = 0 \vee \mathit{lp}_{0,j} = 0  \vee \mathit{lp}_{0,q} = 0 \vee \mathit{lp}_{0,r} = 0    $$
ensures that the outermost loop surrounding the producer is indexed by one of $i$, $j$, $q$, or $r$. Similarly, 
$$\mathit{lp}_{0,i} = 1 \vee \mathit{lp}_{0,j} = 1  \vee \mathit{lp}_{0,q} = 1 \vee \mathit{lp}_{0,r} = 1    $$
guarantees that the second-outermost loop is also indexed by one of the indices of $X_{ i j q r }$. For the fused code structure shown in Fig.~\ref{fig:fused}, we have $\mathit{lp}_{0,r} = 0$ (i.e., the outermost loop for $C_0$ is $r$) and $\mathit{lp}_{0,j} = 1$ (i.e., the second-outermost loop is $j$). Thus, in the fused code, the reference to $X$ will only contain the remaining indices $i$ and $q$, as shown by \texttt{X[q,i]} in Fig.~\ref{fig:fused}.
$\blacksquare$

\paragraph{Consumer constraints}
Next, we create constraints for the consumer contraction $C_j$: the sequence of its outermost $n-l$ loops must match the sequence of the outermost $n-l$ loops for the producer $C_i$. This ensures that the same sequence of $n-l$ loops surround both the producer and the consumer, which is required for fusion that reduces the order of the intermediate from $n$ to $n'$ such that $n' \le n-(n-l)=l$. (In case the constraint solver produces a solution for which more than $n-l$ outermost loops can fused, we can have $n'<l$.) The constraints for $C_j$ include, for each $s$ such that $0 \le s < n-l$ and for each $k \in I_T$, a constraint of the form $$(\mathit{lp}_{i,k}=s) \implies (\mathit{lp}_{j,k}=s)$$ 
For $X_{ i j q r }$ and its consumer $C_1$, we would include constraints connecting $\mathit{lp}_{0,k}$ and $\mathit{lp}_{1,k}$ for each $k\in \{ i, j, q, r \}$ for both  $s=0$ (i.e., the outermost loop) and $s=1$ (i.e., the second-outermost loop).

\paragraph{Statements between producer and consumer}
Finally, we have to consider all assignments that appear between the producer $C_i$ and the consumer $C_j$ in the topological sort order defined by constraint variables $\mathit{ap}_i$ described earlier. For any such assignment, the sequence of the outermost $n-l$ loops that surround it must match the ones for $C_i$ and $C_j$. This is needed in order to have a valid fusion structure. The corresponding constraints are of the following form, for each contraction $C_r$ with $r\ne i$ and $r\ne j$, each $s$ with $0\le s < n-l$, and each $k\in I_T$: $$(\mathit{ap}_i < \mathit{ap}_r < \mathit{ap}_j) \implies ( (\mathit{lp}_{i,k}=s) \implies (\mathit{lp}_{r,k}=s) )$$

\section{Code Generation}

This section details the process of code generation from the constraint system solution.
We describe how to use this solution to generate \emph{concrete index notation}, an IR used by the TACO compiler. This IR is then used by TACO to generate the final C code implementation for the tensor contraction tree. 
The generated C code for the running example is presented in the supplemental materials.

\subsection{Concrete Index Notation}

As discussed in Section~\ref{sec:background}, the Tensor Algebra Compiler (TACO) \cite{kjolstad2017tensor} is a state-of-the-art code generator for sparse tensor computations. While TACO does not address the questions that our work investigates (choice of linear ordering of tensor contractions from a binary contraction tree, selection of fusion structures, and tensor layouts), it does provide code generation functionality for efficient implementations of CSF tensor representations and iteration space traversals. We use concrete index notation~\cite{workspaces2019cgo}, the TACO IR that captures a computation over sparse tensors through a set of computation constructs. 
The two constructs relevant to our work are \texttt{forall} and \texttt{where}. 
A \texttt{forall} construct denotes an iteration over some index. A \texttt{where(C,P)} construct denotes a producer-consumer relationship. Here \texttt{C} represents a computation that consumes a tensor being produced by computation \texttt{P}. This construct allows the use of dense workspaces~\cite{workspaces2019cgo}; as discussed in Sec.~\ref{sec:background}, this is an important optimization in TACO. As an illustration, the concrete index notation we generate from the constraint solution for the running example has the following form:
\begin{verbatim}
    forall(r, forall(j,
    where(forall(k, forall(i, R(j, k, i) = Y(k, i) * D(r, j, k))),
           where(forall(q, forall(k, forall(i, Y(k, i) = X(q, i) * C(r, q, k)))),
            forall(p, forall(q, forall(i, X(q, i) = A(p, q, i) * B(r, j, p))))))))
\end{verbatim}

\subsection{Generating Concrete Index Notation}

The constraint solver's output can be abstracted as a sequence of pairs $\langle A, \pi \rangle$, where $A$ is an assignment 
for a binary contraction and $\pi$ is a permutation of the indices appearing in the assignment. The permutation is defined by the values of constraint variables $\mathit{lp}_{i,k}$ described earlier and denotes the order of surrounding loops for $A$. The indices in a reference to a tensor $T$ in $A$ are ordered based on the values of variables $\mathit{dp}_{T,j}$; thus, they are consistent with the order of indices in $\pi$.
The order in the sequence of pairs is defined by the values of variables $\mathit{ap}_i$ and represents a topological sort order of the contraction tree. For the example discussed in the previous section, the sequence is:

\begin{tabular}{lll}
& & \\
\hspace*{3em} & \texttt{<X[r,j,q,i] = A[p,q,i] * B[r,j,p],} & \texttt{[r,j,p,q,i]>} \\
& \texttt{<Y[r,j,k,i] = X[r,j,q,i] * C[r,q,k],} & \texttt{[r,j,q,k,i]>}\\
& \texttt{<R[j,k,i] = Y[r,j,k,i] * D[r,j,k],} & \texttt{[r,j,k,i]>} \\
& & \\
\end{tabular}

\begin{algorithm*}[t]
	\small
 \DontPrintSemicolon
	\LinesNumbered
\SetKwInOut{Input}{input}
\SetKwInOut{Output}{output}
\SetKwProg{Fn}{function}{:}{}
\SetKwFunction{gen}{generate}
\SetKwFunction{remove}{remove}
\Fn{\gen{$S$}}{
 \Input{sequence $S$ of pairs $\langle A, \pi\rangle$; $A$ is an assignment and $\pi$ is a permutation of $A$'s indices}
 \Output{concrete index notation for $S$}
 
 $L \gets \mathit{empty~list}$ \tcp{$L$ is a sequence of indices and/or assignments} 
 $M \gets \mathit{empty~map}$ \tcp{$M$ is a map from an index to a sequence of $\langle A, \pi\rangle$} 
\For{$k \gets 0$ to $|S|-1$} {  
 $\langle A, \pi\rangle \gets S_k$ \; 
 \uIf{$\pi$ is empty}{
        $L.\mathit{append}(A)$ \;
    } \uElse {
    $i \gets \pi.\mathit{first}()$ \tcp{$i$ is the index of the outermost loop for $A$ at this level} 
    \uIf{$i \ne L.\mathit{last}()$}{
        \tcp{$i$ does not match the last element of $L$ and should be added to $L$}
        $L.\mathit{append}(i)$ \;
        $M.\mathit{put}(i,\mathit{empty~list})$ 
     }
     $M.\mathit{get}(i).\mathit{append}(\langle A, \pi\rangle)$ \;
    }
} 
\uIf{$L.\mathit{length}()==1$}{
\lIf{$L.\mathit{first}()$ is an assignment $A$}{\KwRet $A$ }
\uIf{$L.\mathit{first}()$ is an index $i$}{
 \tcp{single index $i$ in $L$; create a 'forall' construct for $i$ }
    \KwRet \textbf{\texttt{\color{blue} forall(}} $i$ \textbf{\texttt{\color{blue} ,}} \gen(\remove($i$, $M.\mathit{get}(i)$)) \textbf{\texttt{\color{blue} )}}
  }
} \uElse {
\tcp{several indices and/or assignments in $L$; create a 'where' construct } 
    \KwRet \textbf{\texttt{\color{blue} where(}} \gen($M.\mathit{get}(L.\mathit{last}())$) \textbf{\texttt{\color{blue} ,}} \gen($S.\mathit{truncate}(M.\mathit{get}(L.\mathit{last}()))$) \textbf{\texttt{\color{blue} )}}
}}
	\caption{TACO IR Generation\label{algo:generate}}
\end{algorithm*}

Algorithm~\ref{algo:generate} details the process of creating the TACO IR from such an input. Function \texttt{generate} is initially invoked with the entire sequence of pairs $\langle A, \pi \rangle$ based on the constraint system's solution. At each level of recursion, the function processes a sequence $S$ of such pairs.
There are two stages of processing. In the first stage (lines 3--12), a sequence $L$ of assignments and indices is constructed. 
One can think of the elements of $L$ as representing eventual assignments and loops that will be introduced in the TACO IR. For example, an index $i$ in $L$ will eventually lead to the creation of a \texttt{forall(i,...)} construct. Similarly, an assignment in $L$ will produce an equivalent assignment in the TACO IR. Both cases are illustrated below.

During this first stage, for each element $\langle A, \pi \rangle$ of $S$, in order, we need to decide whether the loop structure encoded by $\pi$ can be fused with the loop structure of the previous element of $S$, at this level of loop nesting. 
For example, the sequence shown above contains permutation \texttt{[r,j,p,q,i]} in the first pair of $S$, followed by \texttt{[r,j,q,k,i]} in the second pair. The processing of the first pair will add index \texttt{r} to $L$. In the processing of the second pair, the outermost index \texttt{r} matches the current last element of $L$, and thus \texttt{r} is a common loop for both assignments. The processing of the third pair considers permutation \texttt{[r,j,k,i]}, whose outermost index again matches the last element of $L$. Thus, at the end of the stage, $L$ contains one element: the index \texttt{r}. In a more general case, a combination of indices and assignments could be added to $L$. For example, if the input sequence is 
\texttt{<A0,[i]>}, \texttt{<A1,[]>}, $L$ contains two elements---\texttt{i} followed by \texttt{A1}---which eventually leads to the creation of \texttt{where(A1,forall(i,A0))} as described shortly.

As part of this process, for each index in $L$ the algorithm records the sub-sequence of relevant pairs from $S$. This information is stored in map $M$, with keys being the indices that are recorded in $L$. For the running example, \texttt{r} is mapped in $M$ to the sequence of all three input pairs. This list of pairs is then used in the second stage of processing to generate a construct of the form \texttt{forall(r,...)}.

The second stage (lines 13--18) considers three cases. If $L$ contains a single assignment, this assignment simply becomes the result of IR generation (line 14). If $L$ contains a single index $i$, this index can be used to create a \texttt{forall(i,...)} construct that surrounds all pairs recorded in $M.\mathit{get}(i)$. This creation is shown at line 16. The pairs in $M.\mathit{get}(i)$ are first processed by a helper function \texttt{remove} and then used to recursively generate the body of the \texttt{forall}. The helper function, which is not shown in the algorithm, plays two roles. Both are illustrated by the modified pairs below, which are obtained by calling \texttt{remove(r,}$M.\mathit{get}($\texttt{r}$)$\texttt{)}. 

\begin{tabular}{lll}
& & \\
\hspace*{3em} & \texttt{<X[j,q,i] = A[p,q,i] * B[r,j,p],} & \texttt{[j,p,q,i]>} \\
& \texttt{<Y[j,k,i] = X[j,q,i] * C[r,q,k],} & \texttt{[j,q,k,i]>}\\
& \texttt{<R[j,k,i] = Y[j,k,i] * D[r,j,k],} & \texttt{[j,k,i]>} \\
& & \\
\end{tabular}

First, \texttt{remove} eliminates \texttt{r} from the start of all permutations $\pi$. This reflects the fact that a \texttt{forall(r,...)} is created at line 16. Second, 
the function removes \texttt{r} from all intermediate tensor references for which both the producer and the consumer are in $M.\mathit{get}($\texttt{r}$)$. For example, 
\texttt{X[r,j,q,i]} appears in the first pair (the producer) and in the second pair (the consumer). Both are surrounded by the common loop \texttt{r}, which means that \texttt{X} can be reduced from order-$4$ to order-$3$, and thus the reference is rewritten as \texttt{X[j,q,i]}. A similar change is applied to \texttt{Y[r,j,k,i]}. 

At the next level of recursion, this sequence becomes the input to \texttt{generate}. During that processing, $L$ contains  only index \texttt{j} and \texttt{remove(j,}$M.\mathit{get}($\texttt{j}$)$\texttt{)} is called to obtain the modified sequence 

\begin{tabular}{lll}
& & \\
\hspace*{3em} & \texttt{<X[q,i] = A[p,q,i] * B[r,j,p],} & \texttt{[p,q,i]>} \\
& \texttt{<Y[k,i] = X[q,i] * C[r,q,k],} & \texttt{[q,k,i]>}\\
& \texttt{<R[j,k,i] = Y[k,i] * D[r,j,k],} & \texttt{[k,i]>}\\
& & \\
\end{tabular}

Then \texttt{generate} is called on this sequence. At that level of recursion, $L$ contains three indices: \texttt{p}, \texttt{q}, and \texttt{k}. This illustrates the third case in the processing of $L$. Line 18 shows the creation of a \texttt{where} construct for this case. Since \texttt{k} is the last element of $L$, the first operand of \texttt{where} is the IR generated for the sub-sequence corresponding to \texttt{k}, which here contains a single pair

\begin{tabular}{lll}
& & \\
\hspace*{3em} & \texttt{<R[j,k,i] = Y[k,i] * D[r,j,k],} & \texttt{[k,i]>}\\
& & \\
\end{tabular}

Recall that this first operand of \texttt{where} corresponds to a consumer of a tensor---in this case, tensor \texttt{Y}. The producer of \texttt{Y} appears in the second operand of \texttt{where}, which is generated from the first two pairs from the original sequence: 

\begin{tabular}{lll}
& & \\
\hspace*{3em} & \texttt{<X[q,i] = A[p,q,i] * B[r,j,p],} & \texttt{[p,q,i]>} \\
& \texttt{<Y[k,i] = X[q,i] * C[r,q,k],} & \texttt{[q,k,i]>}\\
& & \\
\end{tabular}

At line 18, $S.\mathit{truncate}$ denotes an operation to produce this desired prefix of $S$ by excluding the sub-sequence defined by $M.\mathit{get}(L.\mathit{last}())$. The IR generated from this prefix itself contains a nested \texttt{where} construct which captures a producer-consumer computation for \texttt{X}. At the end of processing, the resulting overall structure has the form 
\begin{verbatim}
    forall(r, forall(j, where(forall(k, forall(i, A2)), 
                              where(forall(q, forall(k, forall(i, A1))), 
                                    forall(p, forall(q, forall(i, A0)))))))
\end{verbatim}



\section{Experimental Evaluation}
\label{sec:exp}

We evaluate the performance of \ourtool-generated code on several sparse tensor networks.
Section~\ref{subsec:DLPNO} presents a case study of sparse tensor computations arising from recent developments with linear-scaling methods in quantum chemistry~\cite{pinski2015sparse}. 
Section~\ref{subsec:cpd} evaluates performance on the Matricised Tensor Times Khatri-Rao Product (MTTKRP) computation~\cite{kolda2009tensor}.
Section~\ref{subsec:tucker} presents performance on the TTMc (Tensor Times Matrix chain) expression that is the performance bottleneck for the Tucker decomposition algorithm~\cite{kolda2009tensor}.

All experiments were conducted on an AMD Ryzen Threadripper 3990X 64-Core processor with 128 GB RAM.
Reported performance improvements are all for single thread execution.
Optimization flags "-O3 -fast-math" were used to compile the C code, with the GCC 9.4 compiler.

We compare \ourtool\ against state-of-the-art sparse tensor compilers and libraries:

\noindent{\bf TACO}\footnote{TACO code: https://github.com/tensor-compiler/taco}: As discussed in detail earlier, \ourtool\ uses TACO for generation of C code after co-optimization for tensor layout choice, schedule for the contractions, loop fusion, and mode reduction of intermediate tensors. We compare the performance of \ourtool-generated code with that achieved by using direct use of TACO. This was done in two ways: (1) Direct $N$-ary contraction code was generated by TACO, where a single multi-term tensor product expression was provided as input with the same mode order for tensors produced by \ourtool's constraint solver (described in Sec.~\ref{sec:constraints}); (2) TACO was used to generate code for an unfused sequence of binary contractions with the same mode order for tensors as generated by \ourtool. 

\noindent{\bf SparseLNR}\footnote{SparseLNR code: https://github.com/adhithadias/SparseLNR}: SparseLNR takes as input a multi-term tensor product expression and generates fused code for it by transforming it internally to a sequence of binary contractions. We evaluated its performance by providing the same multi-term tensor expression used for comparison with TACO. 
Directly providing a sequence of binary contractions to SparseLNR is not applicable, since it is not designed to generate fused code from such input.

\noindent{\bf Sparta}\footnote{Sparta code: https://github.com/pnnl/HiParTI/tree/sparta}: We used Sparta to compute the sequence of binary tensor contractions produced by \ourtool. However, Sparta's kernel implementation internally requires that the contraction index be at the inner-most mode for one input tensor and at the outer-most mode for the other input tensor. If the provided input tensors do not satisfy this condition, explicit tensor transposition is performed by Sparta before performing the sparse tensor contraction. Since the tensor layout generated by \ourtool\ might not conform to Sparta's constraints, we instead performed an exhaustive study that evaluated all combinations of distinct tensor layout orders that would not need additional transpositions for Sparta. We report the lowest execution time among all evaluated configurations. 
\subsection{Computing Sparse Integral Tensors for DLPNO Methods in Quantum Chemistry}
\label{subsec:DLPNO}


Recent developments in predictive-quality quantum chemistry have sought to reduce their computational complexity from a high-order polynomial in the number of electrons $N$ (e.g., $\mathcal{O}(N^7)$ and higher for predictive-quality methods like coupled-cluster~\cite{bartlett2007coupled}) to linear in $N$, by exploiting various types of sparsity of electronic wave functions and the relevant quantum mechanical operators~\cite{riplinger2016sparse}.




\begin{figure}
\includegraphics[width=0.4\linewidth]{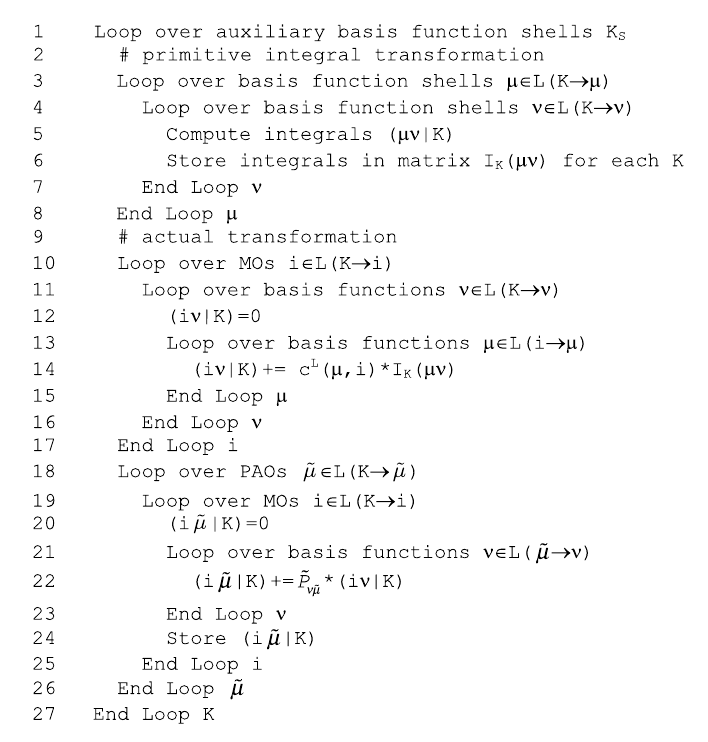} 
\caption{Computation of sparse integral tensors (reproduced from ~\cite{pinski2015sparse})}
\label{fig:pinski_code}
\end{figure}

\begin{figure}[t]
    \begin{subfigure}{0.25\textwidth}
\includegraphics[scale=0.08]{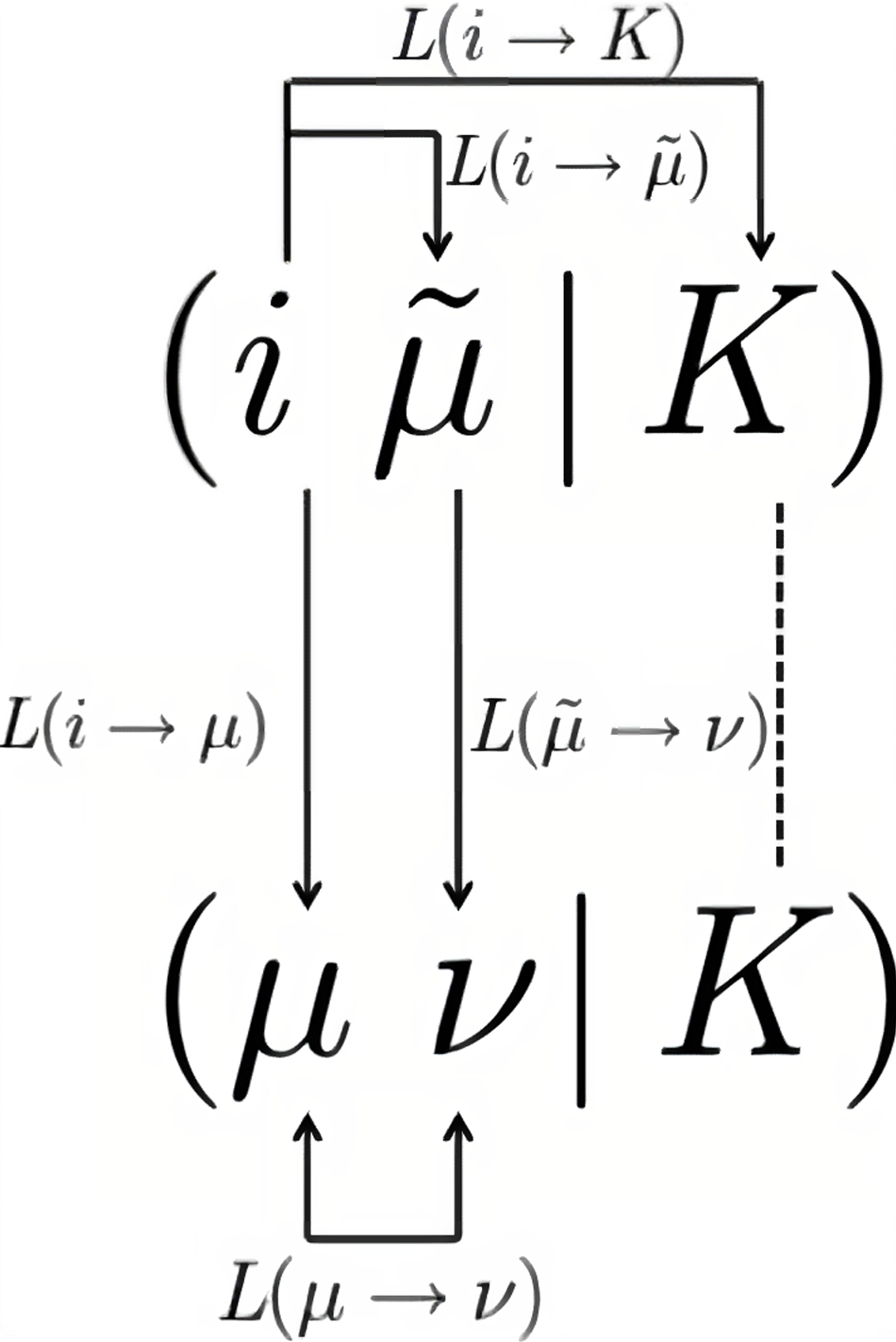}
        \caption{Sparse maps involved in the computation of DLPNO integrals.}
    \label{fig:sparse_maps}    
\end{subfigure}
\hfill
\begin{subfigure}{0.26\textwidth}
\includegraphics[scale=0.22]{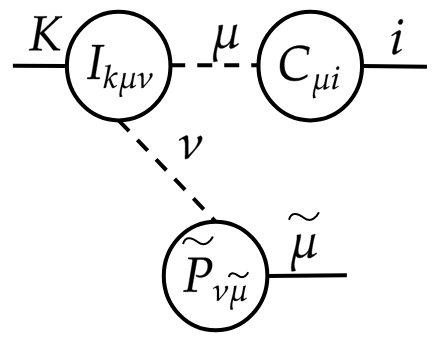}
\caption{Sparse tensor network for unrestricted evaluation of DLPNO integrals [Eq. \eqref{eq:I_3c_unrestricted}].}
\label{fig:mp2_nofilter}
\end{subfigure}
\hspace*{\fill}
\begin{subfigure}{0.25\textwidth}
\hspace*{-1em}\includegraphics[scale=0.20]{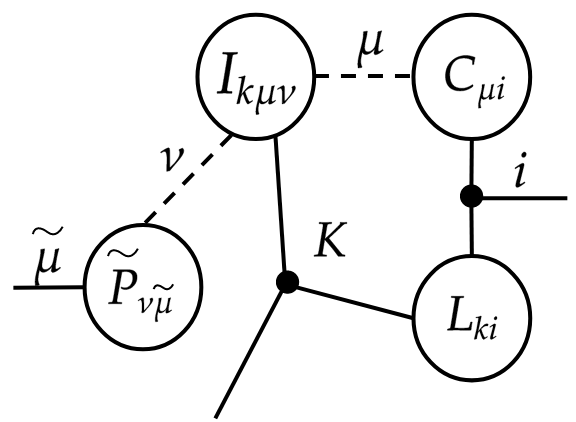}
\caption{Sparse tensor network for 3 centered integral, with sparse tensor $L_{Ki}$ to impose additional sparsity in result tensor [Eq. \eqref{eq:I_3c_restricted}].}
\label{fig:mp2_filter}
\end{subfigure}
\caption{Sparse integral tensor case study}
\label{fig:scheme1}
\end{figure}

The few efficient practical realizations of DLPNO (Domain-based Local Pair Natural Orbital) and other similar methods, e.g., the Orca package \cite{neese2020orca}, have developed custom implementations of sparse tensor algebra,  without any utilization of generic infrastructure for sparse tensor computations. In this section, we present a case study that demonstrates the potential for using \ourtool\ to automatically generate code that can address the kinds of sparsity constraints that arise in the implementation of DLPNO and similar sparse formulations in quantum chemistry.

A key step in the DLPNO methods is the evaluation of matrix elements (integrals) of the electron repulsion operator that was first formulated in a linear-scaling fashion by Pinski et al.~\cite{pinski2015sparse}. 
The first key step of the DLPNO integral evaluation involves a multi-term tensor product of 3 sparse tensors (Fig.~\ref{fig:mp2_nofilter} shows a sparse tensor network corresponding to the expression):
\begin{align}
    E_{K i \tilde{\mu}} = I_{K\mu\nu} \times C_{\mu i} \times \tilde{P}_{\nu \tilde{\mu}}
    \label{eq:I_3c_unrestricted}
\end{align}
The indices of the three input tensors and output tensor correspond to four pertinent spaces, ordered from least to most numerous: (1) localized {\em molecular orbitals} (MO; indexed in the code by $i$), (2) {\em atomic orbitals} (AO; indexed by $\mu$ and $\nu$), (3) {\em projected atomic orbitals}~\cite{pulay1983localizability} (PAO; indexed by $\tilde{\mu}$), and (4) density fitting atomic orbitals (DFAO; indexed by $K$).

Fig.~\ref{fig:pinski_code} shows pseudocode for its computation in the Orca quantum chemistry package \cite{neese2020orca} as a sequence of 3 stages: (1) form  integral $I_{K\mu\nu}$ (lines 3--8; denoted by $(\mu \nu|K)$); (2) compute intermediate tensor $(i \nu | K)$ as the product $(\mu \nu|K) \times C_{\mu i}$ (lines 10--17; $C_{\mu i}$ is denoted $C^L(\mu,i)$); (3) compute final result $E_{K i \tilde{\mu}}$ (denoted $(i \tilde{\mu} | K) $) as the tensor product $(i \nu | K) \times \tilde{P}_{\nu \tilde{\mu}}$ (lines 18--26).

The ranges of loops in the code are governed by various sparsity relationships or \emph{sparse maps} between pairs of index spaces, as illustrated in Fig.~\ref{fig:sparse_maps} (reproduced from Pinski et al. \cite{pinski2015sparse}). A sparse map associates a subset of elements in the range space for each element in the domain and inverse maps exist for each map. Sparse maps are used in the code in Fig.~\ref{fig:pinski_code} to reduce computations. For example, for each $K$, the intermediate tensor $(i \nu | K)$ obtained by contracting $(\mu \nu|K) \times C_{\mu i}$ would have nonzero $i$ corresponding to the union of nonzeros in $C_{\mu i}$  across all nonzero $\mu$ in $(\mu \nu|K)$. However, as seen in line 10 of Fig.~\ref{fig:pinski_code}, the range of $i$ is restricted by an available pre-computed sparse map $L(K \rightarrow i)$. This enables a reduction of the executed operations and only a subset of all elements of this tensor network are evaluated. Fig.~\ref{fig:mp2_filter} shows a 4-term sparse tensor network where an additional 0/1 sparse matrix $L_{Ki}$ has been added to the base tensor network in Fig.~\ref{fig:mp2_nofilter}, corresponding to the known sparse map $L(K \rightarrow i)$. This can equivalently be expressed as a multi-term tensor product expression:
\begin{align}
    E_{K i \tilde{\mu}} = I_{K\mu\nu} \times C_{\mu i} \times \tilde{P}_{\nu \tilde{\mu}} \times L_{Ki}
    \label{eq:I_3c_restricted}
\end{align}
The inclusion of such sparse maps as additional nodes in the base tensor network has the same beneficial effect of reducing computations as the manually implemented restriction in the loop code of Fig.~\ref{fig:pinski_code}.
In our experimental evaluation, we evaluate both forms of the sparse tensor networks in Fig.~\ref{fig:scheme1}, representing the \emph{unrestricted} form (Eq.~\ref{eq:I_3c_unrestricted}, Fig.~\ref{fig:mp2_nofilter}) and the \emph{restricted} form (Eq.~\ref{eq:I_3c_restricted}, Fig.~\ref{fig:mp2_filter}).

We computed the DLPNO integrals for 2-dimensional solid helium lattices with the geometry described in \cite{helium_paper}.
This computation was done for a $5\times 5$ lattice of 25 atoms and a $10 \times 10$ lattice of 100 atoms,
orbital and density fitting basis sets 6-311G~\cite{g09} and the spherical subset of def2-QZVPPD-RIFIT~\cite{qzvpp, def2svpd}, and cc-pVDZ-RIFIT~\cite{ccpvdz, ccpvdzrifit}.
All quantum chemistry data was prepared using the Massively Parallel Quantum Chemistry package \cite{mpqc}.

\begin{figure}[t]
\begin{subfigure}{0.55\textwidth}
    \includegraphics[scale=0.21]{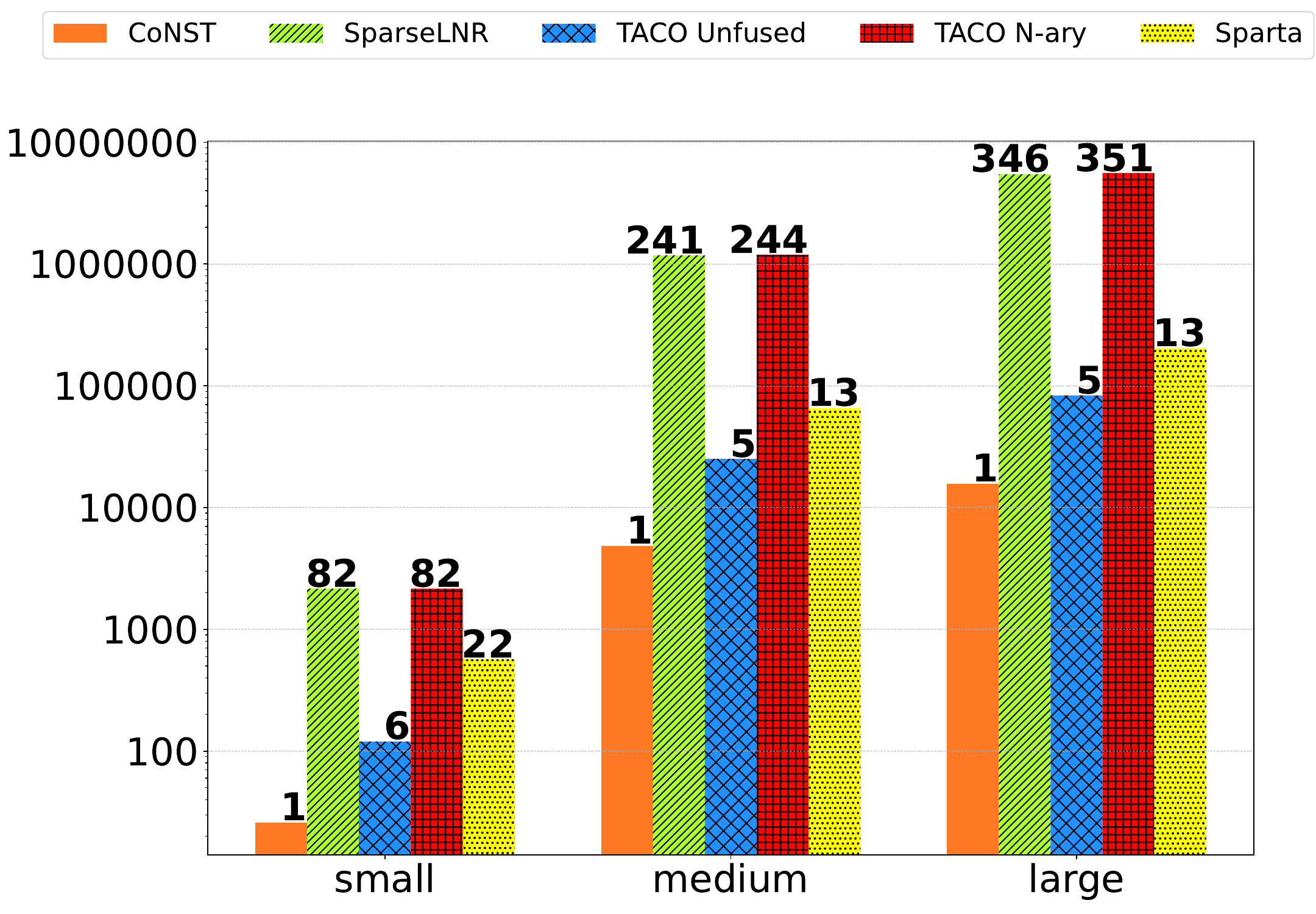}
    \caption{}
    \label{fig:nofilter_3c}
\end{subfigure}
\begin{subfigure}{0.40\textwidth}
    \includegraphics[scale=0.21]{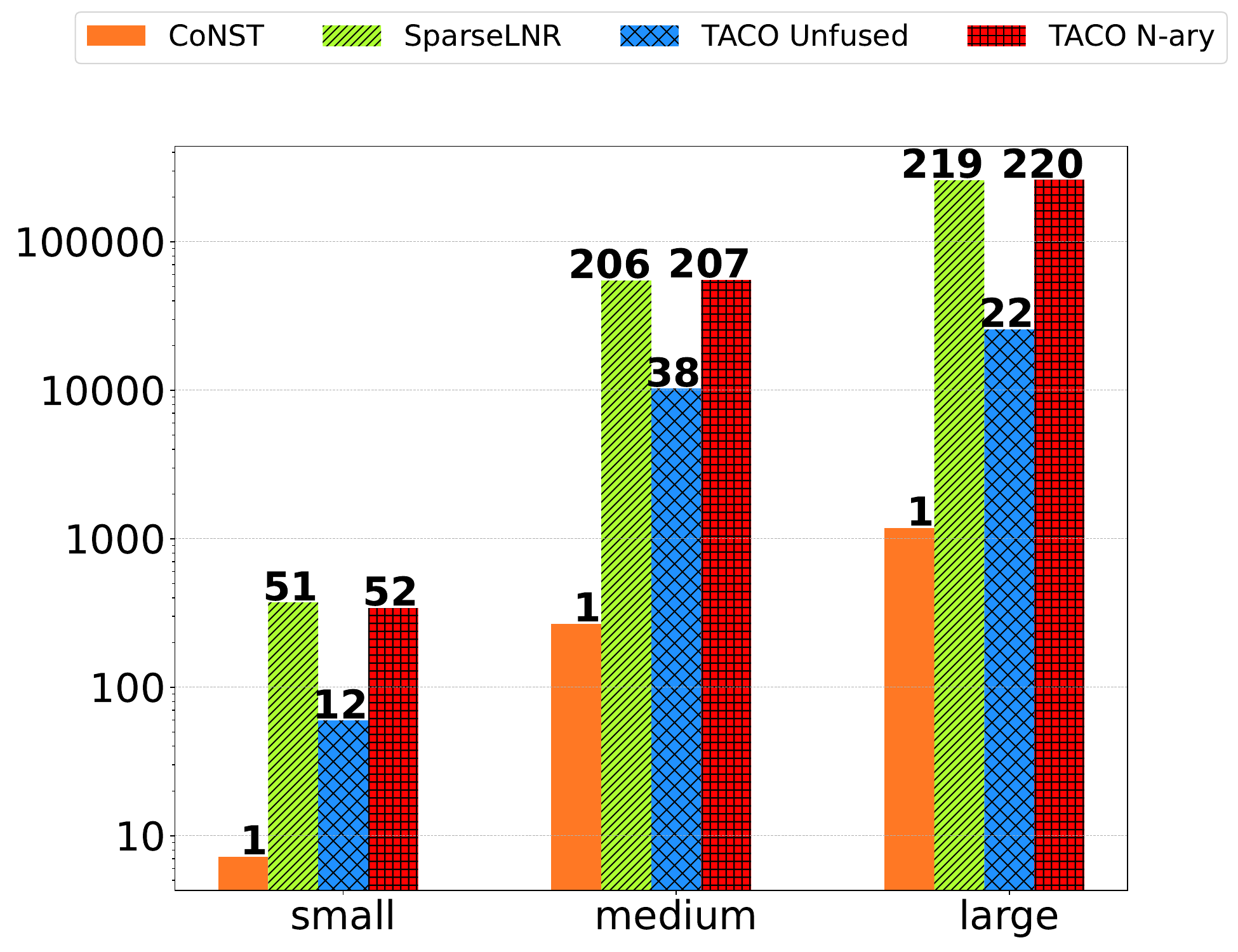}
    \caption{}
    \label{fig:filter_3c}
\end{subfigure}
\caption{Execution time (ms) for evaluation of  3-index integrals (lower is better; Y-axis is in logarithmic scale) using (a) unrestricted [Eq. \eqref{eq:I_3c_unrestricted}] and (b) restricted [Eq.~\eqref{eq:I_3c_restricted}] tensor networks, respectively. See text for the description of ``small'', ``medium'' and ``large'' datasets. The numbers above the bars represent the slowdown of other schemes relative to CoNST.}
\label{fig:3c_int}
\end{figure}

Figure~\ref{fig:nofilter_3c} presents performance data for evaluation of the transformed 3-index integral $E_{K i \tilde{\mu}}$ via Eq.~\ref{eq:I_3c_unrestricted}. It my be seen that \ourtool-generated code is about two orders of magnitude faster than the N-ary code generated by TACO as well as the SparseLNR (for this case SparseLNR was unable to perform loop fusion and simply lowered the input to TACO). TACO-Unfused is much faster than N-ary, due of the sequence of tensor contractions and mode layouts for sparse tensors generated by \ourtool, but it is still about 5-6$\times$ slower than the code generated by \ourtool. The best of the comprehensively evaluated versions for Sparta is also about an order of magnitude slower than \ourtool's code.


The performance data for evaluation $E_{K i \tilde{\mu}}$ using Eq.~\ref{eq:I_3c_restricted} is presented in Figure~\ref{fig:filter_3c}.
Significant speedups can be seen between the execution times in Figure~\ref{fig:nofilter_3c} and Figure~\ref{fig:filter_3c} (the Y-axis scales are different) by use of the additional tensor $L_{Ki}$ for \ourtool, SparseLNR, and TACO N-ary, with the speedup with use of \ourtool\ being roughly the same. However, TACO-Unfused does not improve as much, causing its slowdown with respect to \ourtool\ to get worse. No data for Sparta is presented in Figure~\ref{fig:filter_3c} because of a constraint of Sparta that a tensor product must have a contraction index, which is not he case for the tensor product with $L$.

\begin{wrapfigure}{l}{0.4\textwidth}
    \includegraphics[scale=0.16]{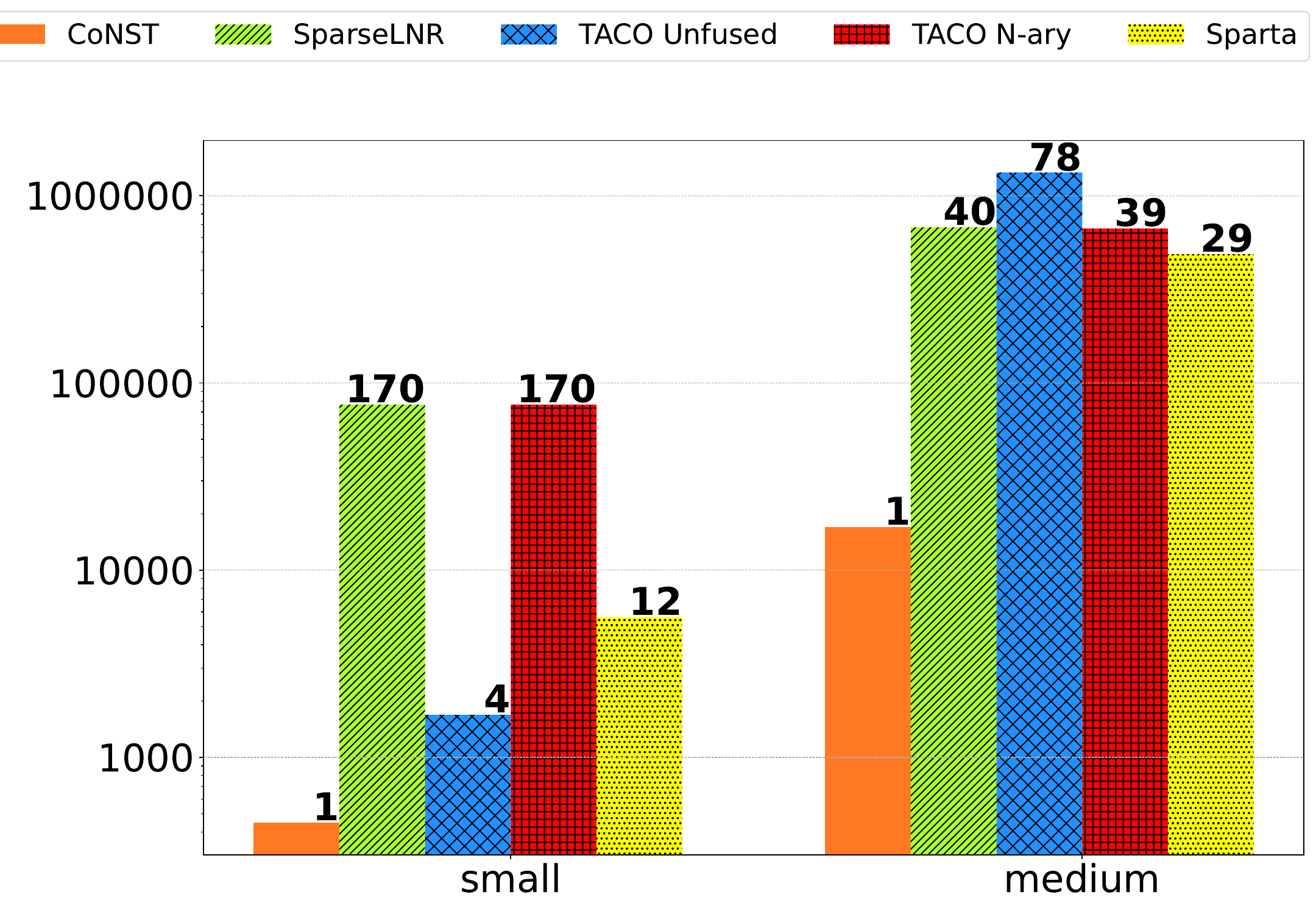}
    \caption{Execution time (ms) of evaluation of 4-index integral via Eq.~\eqref{eq:I_4c_unrestricted} (lower is better). Numbers at the top of the bar are relative execution time (slowdown) with \ourtool\ as 1.} 
    \label{fig:4c_int_res}
\end{wrapfigure}

A subsequent step after formation of the 3-centered integrals is to use them to construct 4-index integrals in DLPNO methods [see Eq. (16) in Ref.~\citenum{pinski2015sparse}] by the following binary contraction:
\begin{align}
    V_{ij\tilde{\mu}\tilde{\nu}} = E_{K i \tilde{\mu}} \times E_{K j \tilde{\nu}},
    \label{eq:I_4c_unrestricted}
\end{align}
using the 3-index input tensor $E$ obtained via Eq.~\eqref{eq:I_3c_unrestricted}.
Performance results are reported in Fig.~\ref{fig:4c_int_res}. \ourtool\ again achieves significant speedup over the alternatives. For this experiment, we could not use the large dataset because of physical memory limitations on our target platform. 

\subsection{Sparse Tensor Network for CP Decomposition}
\label{subsec:cpd}
CP (Canonical Polyadic) Decomposition factorizes a sparse tensor $T$ with $n$ modes into a product of $n$ 2D matrices. For example, a 3D tensor $T_{ijk}$ is decomposed into three dense rank-$r$ matrices $A_{ir}$, $B_{jr}$, and $C_{kr}$. The CP decomposition of a sparse tensor is generally performed using an iterative algorithm that requires $n$ MTTKRP (Matricized Tensor Times Khatri-Rao Product) operations \cite{kolda2009tensor}. For a 3D tensor, the three MTTKRP operations are as follows:

\begin{description}
    \item [] $A'_{ir} = T_{ijk} \times B_{jr} \times C_{kr}$ \hfill $B'_{jr} = T_{ijk} \times A_{ir} \times C_{kr}$ \hfill $C'_{kr} = T_{ijk} \times A_{ir} \times  B_{jr}$
\end{description}

\begin{figure}[t]
    \begin{subfigure}{0.5\textwidth}
    \includegraphics[scale=0.22]{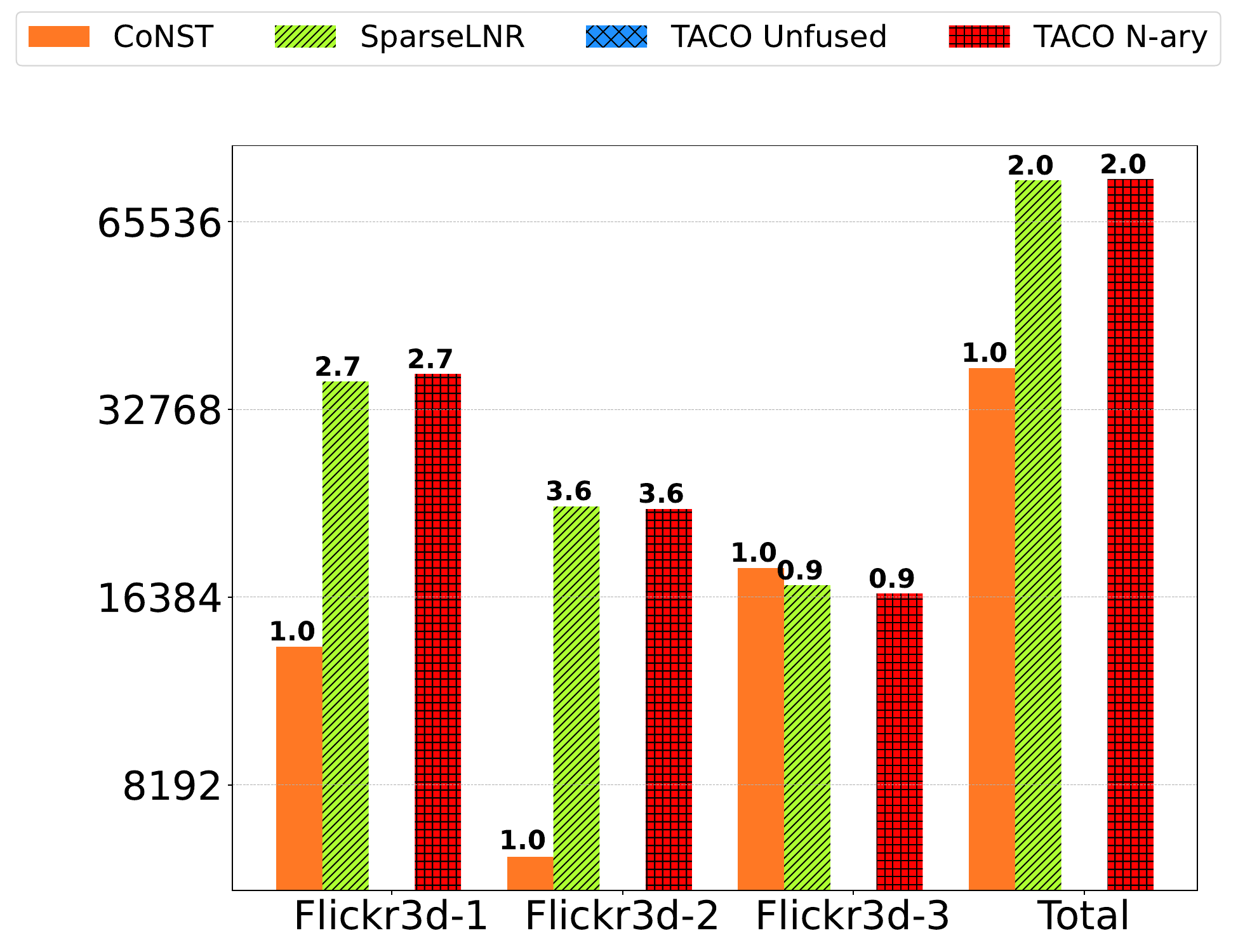}
    \end{subfigure}
\hspace*{\fill}
    \begin{subfigure}{0.46\textwidth}
    \includegraphics[scale=0.22]{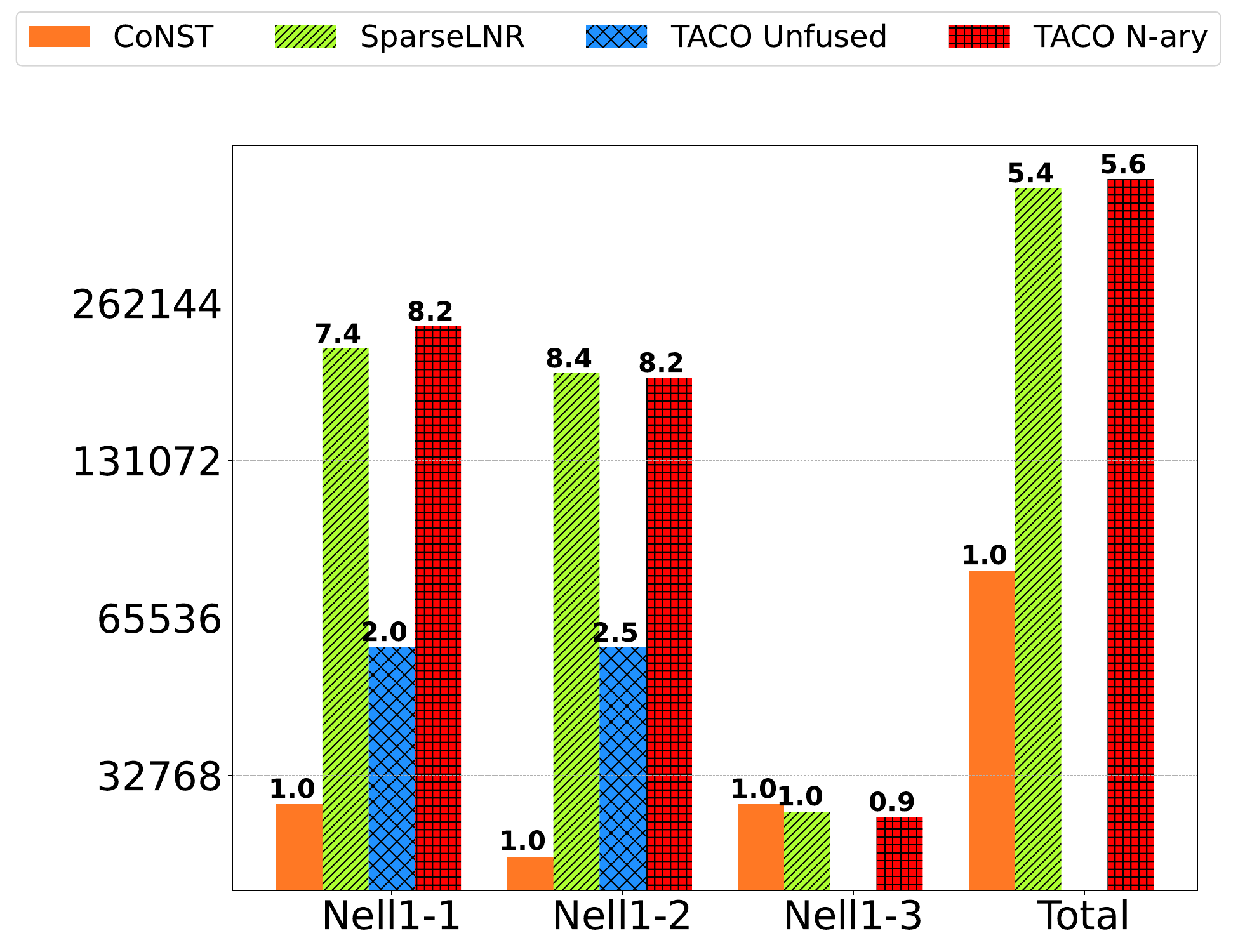}
    \end{subfigure}
\hspace*{\fill}
    \begin{subfigure}{0.5\textwidth}
    \includegraphics[scale=0.21]{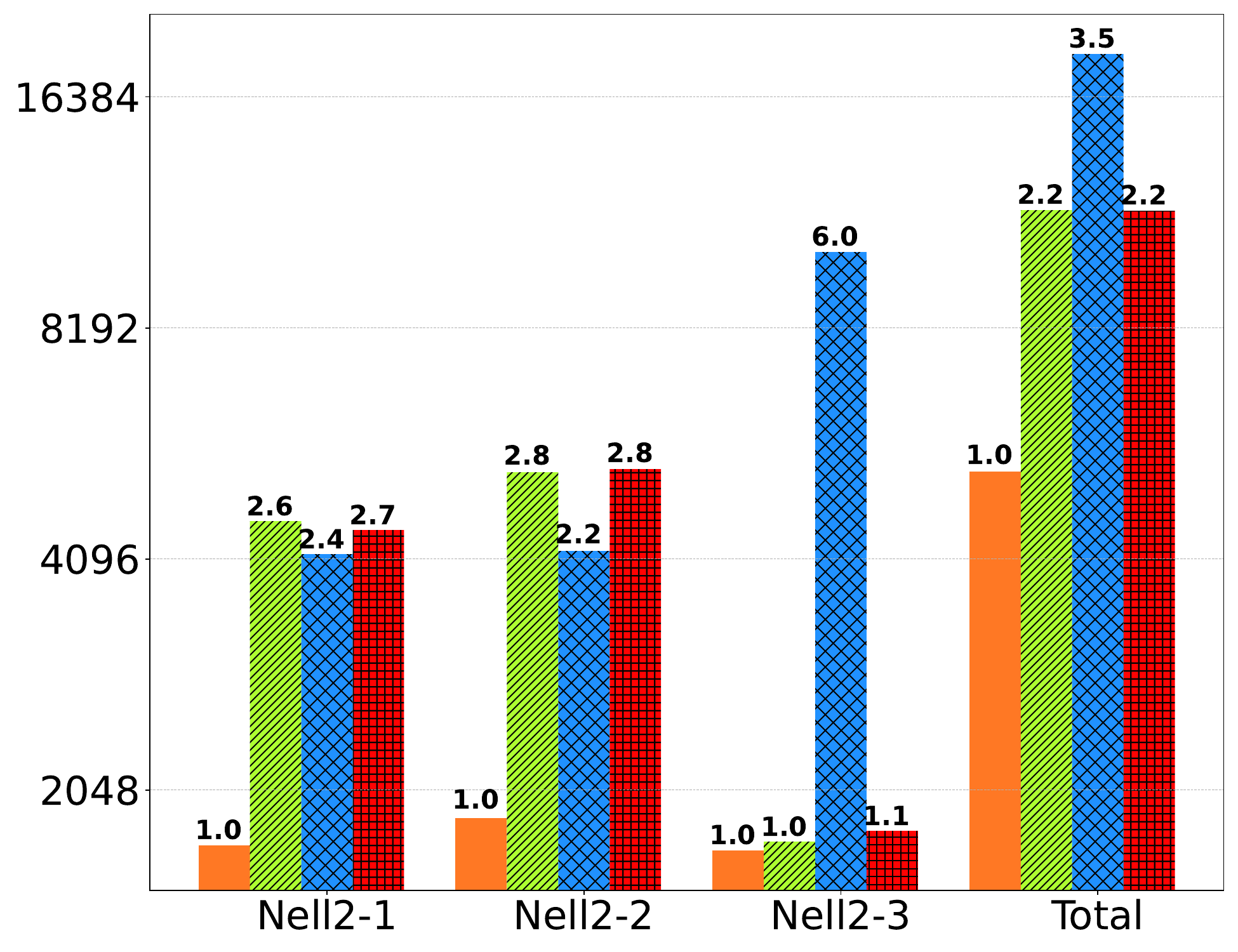}
    \end{subfigure}
\hspace*{\fill}
    \begin{subfigure}{0.45\textwidth}
    \includegraphics[scale=0.21]{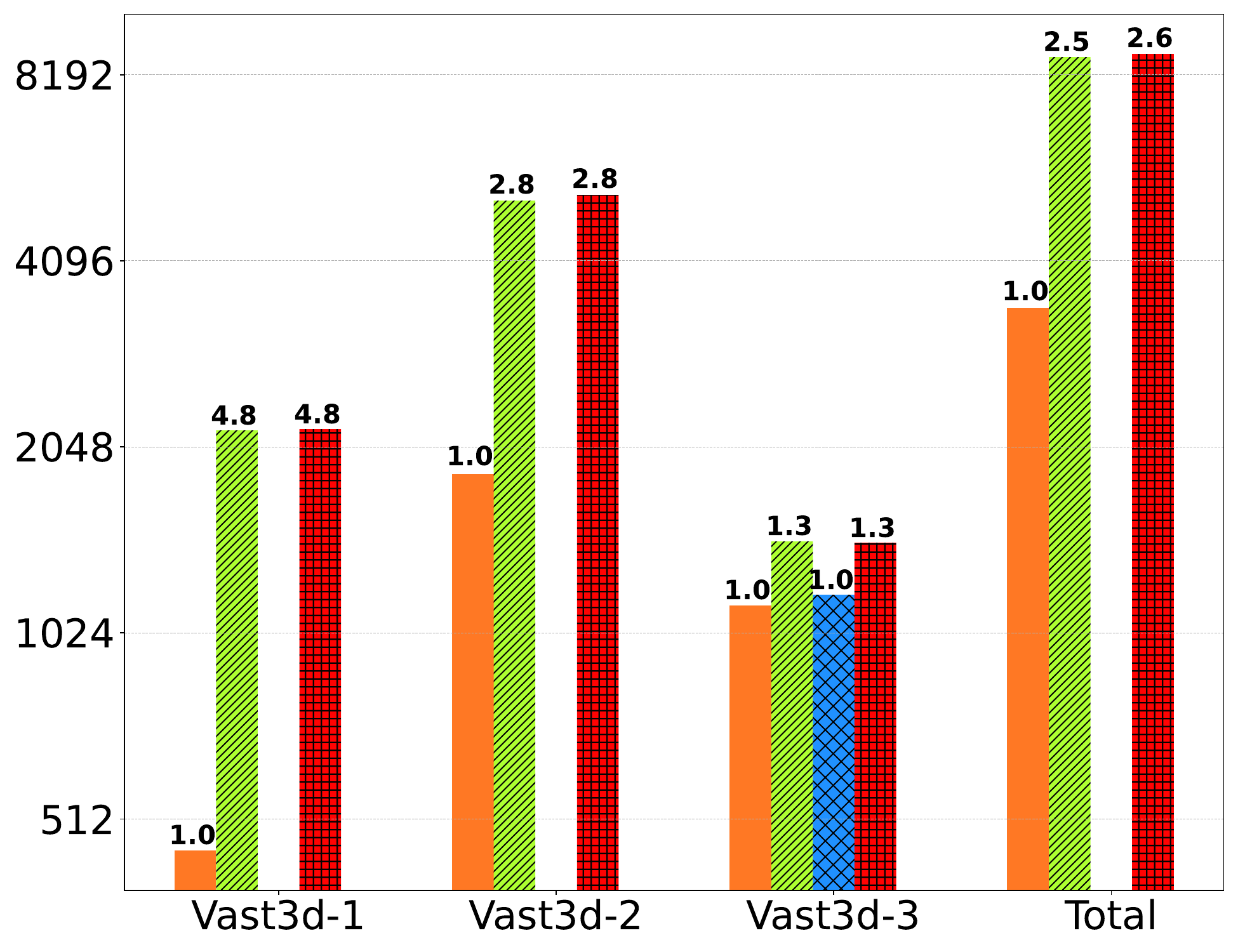}
    \end{subfigure}
\hspace*{\fill}
\caption{Execution time (ms) for MTTKRP operations on the FROSTT tensors. Relative slowdown compared to CoNST is indicated above each bar. Missing bars mean out-of-memory failure (for TACO-Unfused).}
\label{fig:cpd_res}
\end{figure}

Figure~\ref{fig:cpd_res} shows performance for MTTKRP operations for each of the three modes for sparse tensors from the FROSTT benchmark suite \cite{FROSTT}. We used the same four sparse tensors (Flickr3d, Nell1, Nell2, and Vast3d) used in the experimental evaluation of SparseLNR \cite{dias2022sparselnr}.
The rank of factor matrices was set to 50. The time to perform the MTTKRP operation for the three modes varies quite significantly. This is in part due to the the highly non-uniform extents of the three modes for the tensors (as seen in Table~\ref{tab:frostt}) and the asymmetry with respect to the  matrices: each of the three MTTKRP operations for CPD has a different matrix in the output, with the remaining two matrices appearing in the right-hand side of the multi-term tensor product.
For the MTTKRP expression, SparseLNR was not able to perform its \emph{loopFusionOverFission} transformation, so that the code and performance is essentially identical to TACO N-ary.
Considering \ourtool, unlike the case with the previously discussed DLPNO benchmark (Sec.~\ref{subsec:DLPNO}), the \ourtool-generated code is not always faster than the other cases. For each benchmark, for the first two out of the three MTTKRPs \ourtool\ achieves a minimum speedup between 2.0$\times$ and 4.8$\times$ over other schemes, but relative performance is low for the third MTTKRP, ranging between 0.9$\times$ and 1.0$\times$ over the best alternative. However, when considering the total time for all three MTTKRPs needed in each iteration in the iterative algorithm for CP Decomposition, \ourtool\ achieves a minimum speedup of 2$\times$ over all others, across the four benchmarks. Sparta times are not reported for this benchmark because it could not be used: it does not handle tensor contractions with ``batch'' indices that occur in both input tensors and output tensor, as occurs with the second tensor contraction in the binarized sequence for each MTTKRP.

In many cases, creating a sparse intermediate after binarization speeds up the MTTKRP operation.
Therefore, TACO-Unfused outperforms TACO N-ary.
However, for Flickr3d and first two modes of Vast3d, this sparse intermediate is too large to fit in the machine RAM, and TACO-Unfused ends with out-of-memory error.
\subsection{Sparse Tensor Network for Tucker Decomposition}
\label{subsec:tucker}

Tucker decomposition factorizes a sparse tensor $T$ with $n$ modes into a product of $n$ 2D matrices and a dense \emph{core} $n$-mode tensor. For example, a 3D tensor $T_{ijk}$ is decomposed into three rank-$r$ matrices $A_{ix}$, $B_{jy}$, $C_{kz}$, and core tensor $G_{xyz}$. The Tucker decomposition of a sparse tensor is generally performed using the HOOI (High Order Orthogonal Iteration) iterative algorithm that requires $n$ TTMc (Tensor Times Matrix chain) operations \cite{kolda2009tensor}. For a 3D tensor, the three TTMc operations are as follows:

\begin{description}
    \item [] $A'_{iyz} = T_{ijk} \times B_{jy} \times C_{kz}$ \hfill $B'_{jxz} = T_{ijk} \times A_{ix} \times C_{kz}$ \hfill $C'_{kxy} = T_{ijk} \times A_{irxr} \times  B_{jryr}$
\end{description}

\begin{table}[t]
\begin{tabular}{llllll}
\hline
Tensor           & \multicolumn{3}{l}{Dimensions}           & \multicolumn{2}{l}{NNZs}    \\ \hline
flickr-3d        & 320K                     & 2.82M & 1.6M  & \multicolumn{2}{l}{112.89M} \\
nell-2           & 12K                      & 9K    & 288K  & \multicolumn{2}{l}{76.88M}  \\
nell-1           & \multicolumn{1}{c}{2.9M} & 2.14M & 25.5M & \multicolumn{2}{l}{143.6M}  \\
vast-2015-mc1-3d & 165K                     & 11K   & 2     & \multicolumn{2}{l}{26.02M}  \\ \hline
\end{tabular}
\caption{FROSTT tensors and their shapes}
\label{tab:frostt}
\end{table}
\begin{figure}[t]
    \begin{subfigure}{0.30\textwidth}
    \includegraphics[scale=0.18]{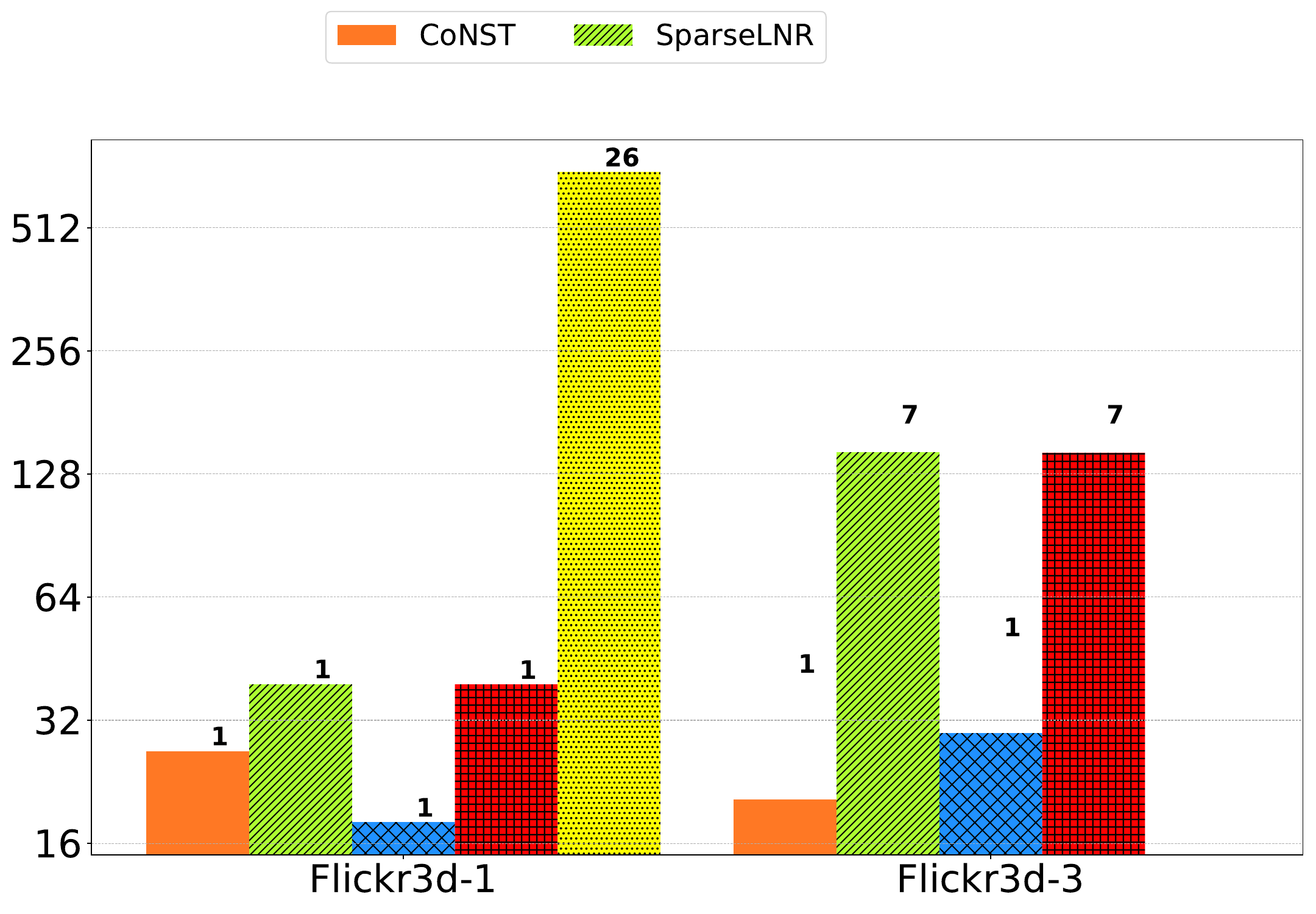}
    \end{subfigure}
\hspace*{\fill}   
    \begin{subfigure}{0.52\textwidth}
    \includegraphics[scale=0.18]{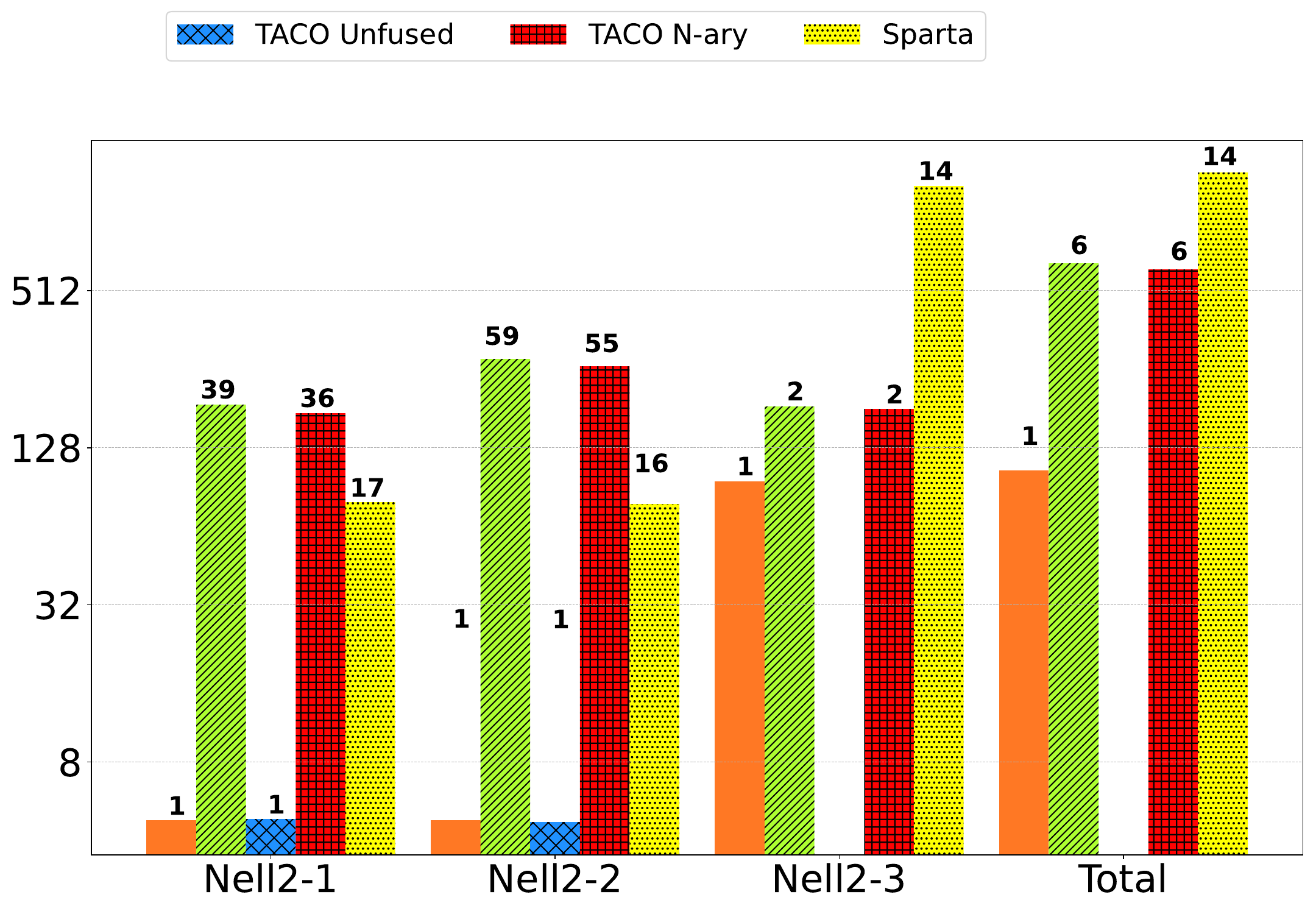}
    \end{subfigure}
    \begin{subfigure}{0.5\textwidth}
    \includegraphics[scale=0.18]{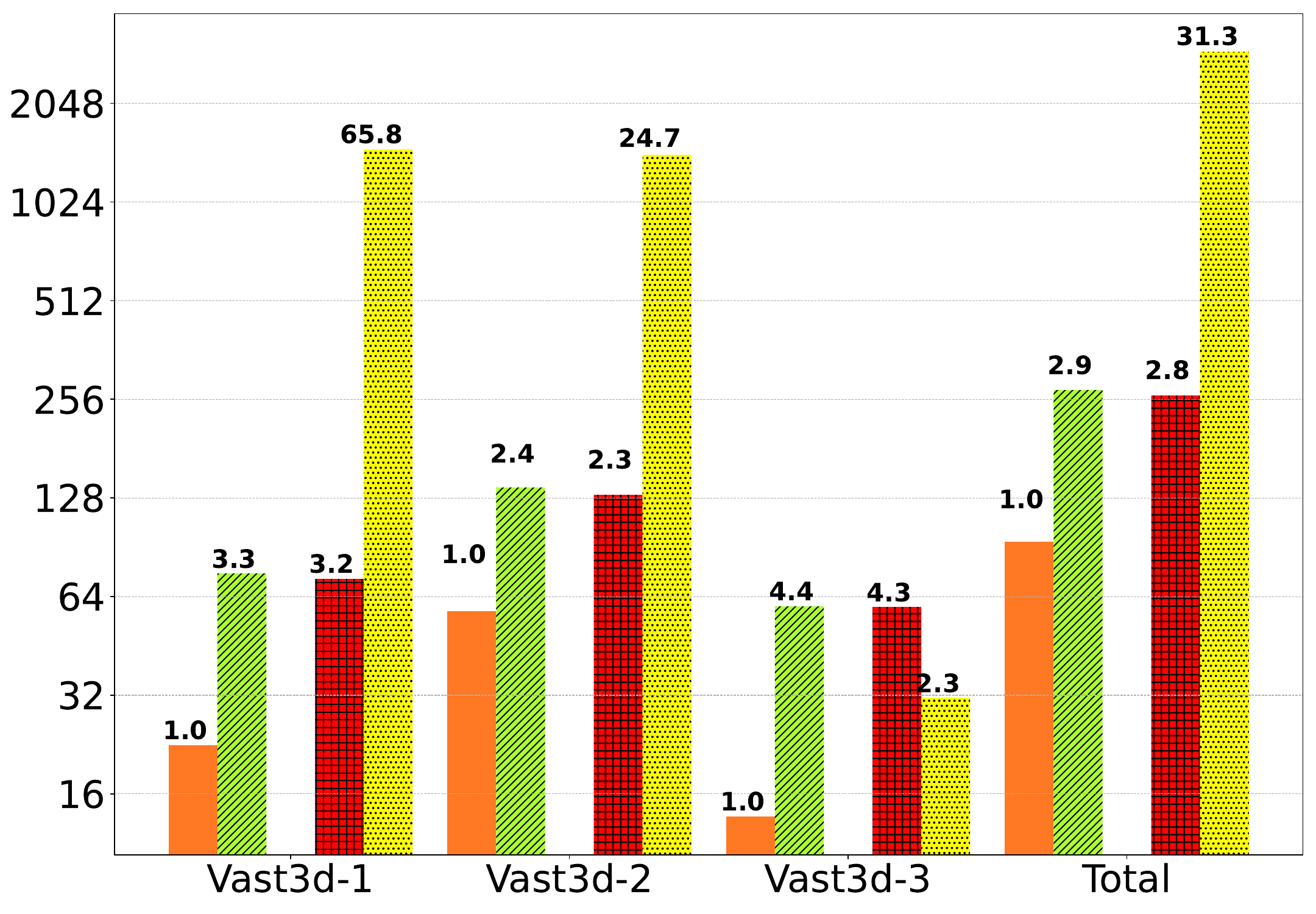}
    \end{subfigure}
    \begin{subfigure}{0.40\textwidth}
    \includegraphics[scale=0.175]{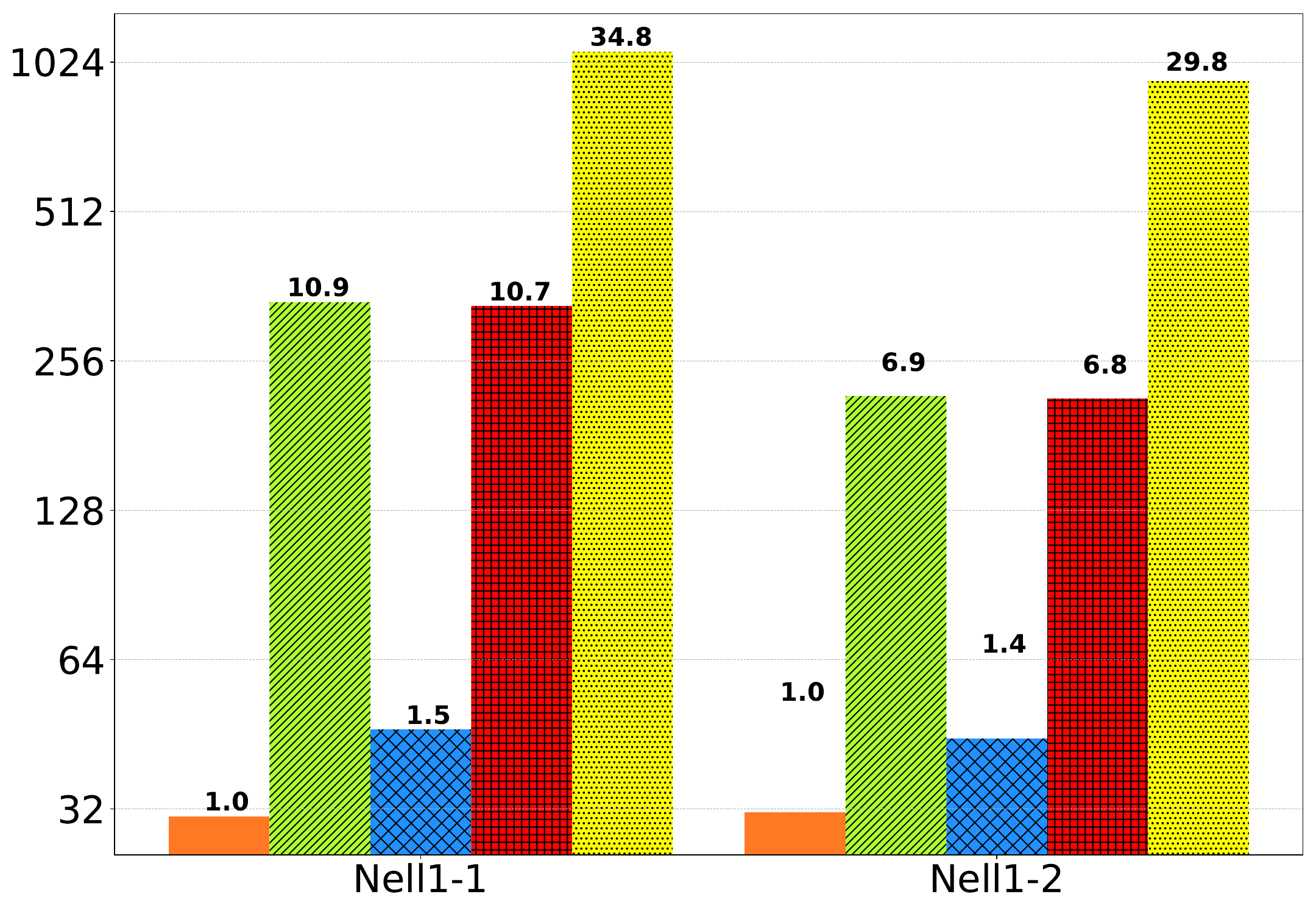}
    \end{subfigure}
\hspace*{\fill}
\caption{Execution time (ms) for TTMc operations on the FROSTT tensors. Relative slowdown compared to CoNST is shown above the bar. Missing bars indicate out-of-memory failure.}
\label{fig:tucker_res}
\end{figure}
\label{subsec:tucker}

Fig.~\ref{fig:tucker_res} presents execution times for the alternative schemes on the four FROSTT tensors.
The mode-2 contraction for Flickr3d and mode-3 contraction for Nell-1 tensor ran out of memory for all methods on 128GB RAM.
TACO-Unfused and Sparta ran out of memory for a larger set of runs because they form high dimensional sparse intermediates in memory.
The rank of decomposition was 16 for Nell-1 and Flickr-3d tensors, and 50 for Vast-3d and Nell-2 tensors.
For the TTMc operation, SparseLNR is not able to perform its \emph{loopFusionOverFission} transformation, so that performance is identical to TACO N-ary.
Sparta runs a flattened matrix-times-matrix operation for a general tensor contraction, and uses a hashmap to accumulate rows of the result.
Since the matrix being multiplied is dense, the hashmap simply adds an overhead. Overall, \ourtool\ generates code that achieves significant speedups over the compared alternatives. 
\label{subsec:tucker}

\section{Related Work}

A comparison between \ourtool\ and the three most related prior efforts was presented in Sec.~\ref{sec:background} and summarized in Table~\ref{tab:comparison_SOA}.
As shown in the previous section, significant performance improvements can be achieved by the code generated through \ourtool's integrated treatment of contraction/loop/mode order for fused execution of general contraction trees, compared to (1) code directly generated by TACO~\cite{kjolstad2017tensor}, (2) fused loop code generated by SparseLNR~\cite{dias2022sparselnr}, and (3) calls to the Sparta library \cite{liu2021sparta} for sparse tensor contractions.

Cheshmi et al. developed sparse fusion~\cite{cheshmi2023runtime},  an inspector-executor strategy for iteration
composition and ordering for fused execution of two sparse kernels. Their work optimizes sparse kernels with loop-carried dependences using runtime techniques. In contrast, our work considers compile-time code generation for a general tree of sparse tensor contractions, where each contraction does not have loop-carried dependences. Tensor mode layout and its interactions with iteration order and mode reduction of intermediate sparse tensors are not considered by Cheshmi et al.~\cite{cheshmi2023runtime}

Work on the sparse polyhedral framework~\cite{sparse-polyhedral} defines general inspector-executor techniques for optimization of sparse computations, e.g., through combinations of run-time iteration/data reordering.
Our approach does not consider run-time inspection/optimization, but rather explores statically the space of possible loop structures and mode orders using a constraint-based formulation. The sparse polyhedral framework has been applied to individual tensor contractions~\cite{zhao2022polyhedral} where tensors are represented in a variety of formats and co-iteration code is derived using polyhedral scanning. Their approach does not consider fusion or reordering of loops/tensor modes, but does provide general reasoning and optimization of individual contractions. 



SparseTIR~\cite{ye2023sparsetir} is an approach to represent sparse tensors in composable formats and to enable program transformations in a composable manner. 
The sparse compilation support in the MLIR infrastructure \cite{sparse_mlir} enables integration of sparse tensors and computations with other elements of MLIR, as well as TACO-like code generation. SpTTN-Cyclops~\cite{kanakagiri2023minimum} is an extension of CTF (Cyclops Tensor Framework) \cite{ctf} to optimize a sub-class of sparse tensor networks. In contrast to \ourtool, which can handle arbitrary sparse tensor networks, SpTTN-Cyclops only targets a product of a single sparse tensor with a network of several dense tensors. Indexed Streams~\cite{kovach2023indexed} 
develops a formal operational model and intermediate representation for fused execution of tensor contractions, using both sparse tensor algebra and relational algebra, along with a compiler to generate code.
Tian et al.~\cite{tian2021high} introduce a DSL to support dense and sparse tensor algebra algorithms and sparse tensor storage formats in the COMET compiler~\cite{mutlu2020comet}, which generates code for a given tensor expression. None of these efforts address the coupled optimization of tensor layout, contraction schedule, and mode reduction for intermediates in fused code being performed by \ourtool.

\section{Conclusions}

Effective fused code generation for sparse tensor networks depends on several inter-related factors: schedule of binary contractions, permutation of nested loops, and layout order of tensor modes. We demonstrate that an integrated constraint-based formulation can capture these factors and their relationships, and can produce fused loop structures for efficient execution. Our experimental evaluation confirms that this approach significantly advances the state of the art in achieving high performance for sparse tensor networks. An important next step is the parallelization of the generated code for multicore processors and GPUs and use of the developed framework to generate high-performance implementations for sparse tensor networks needed by computational scientists (e.g., in quantum chemistry). 

\bibliographystyle{ACM-Reference-Format}
\bibliography{refs}

\end{document}